\documentclass[%
 reprint,
 superscriptaddress,
 showpacs,
 amsmath,amssymb,
 aps,
 prd,
 showkeys,
 groupedaddress,
]{revtex4-1}

\usepackage{booktabs,graphicx,mathrsfs,verbatim,amsmath,units}
\usepackage{graphicx}
\usepackage{dcolumn}
\usepackage{bm}
\usepackage{mhchem}
\usepackage{siunitx}
\usepackage{multirow}
\usepackage[none]{hyphenat}

\newcommand\figref[1]{Fig.\,\ref{#1}}
\newcommand\eqnref[1]{Eq.\,\ref{#1}}

\newcommand\citeref[1]{Ref.\,\cite{#1}}

\begin{document}

\newcommand{\columbia}{\affiliation{Physics Department, Columbia University, New York, NY, USA}}

\preprint{APS/123-QED}

\title{Simultaneous measurement of the light and charge response of liquid xenon to low-energy nuclear recoils at multiple electric fields}

\author{E.~Aprile}
\author{M.~Anthony}\email[]{mda2149@columbia.edu}
\author{Q.~Lin}\email[]{ql2265@columbia.edu}
\author{Z.~Greene}
\author{P.~de~Perio}
\author{F.~Gao}
\author{J.~Howlett}
\author{G.~Plante}
\author{Y.~Zhang}
\author{T.~Zhu}

\columbia
\date{\today}

\begin{abstract}
Dual-phase liquid xenon (LXe) detectors lead the direct search for particle dark matter.
Understanding the signal production process of nuclear recoils in LXe is essential for the interpretation of LXe based dark matter searches. 
Up to now, only two experiments have simultaneously measured both the light and charge yield at different electric fields, neither of which attempted to evaluate the processes leading to light and charge production.  
In this paper, results from a neutron calibration of liquid xenon with simultaneous light and charge detection are presented for nuclear recoil energies from 3--74 \,keV, at electric fields of 0.19, 0.49, and 1.02\,kV/cm.
No significant field dependence of the yields is observed.
\end{abstract}

 

\maketitle



\section{Introduction} 
Liquid xenon (LXe) time projection chambers (TPCs) lead the search for weakly interacting massive particles (WIMPs) with detectors of increasing size and decreasing background \cite{aprile2017first, cui2017dark, akerib2016improved, aprile2016xenon100}.
Since WIMPs are hypothesized to interact primarily with atomic nuclei and the differential scattering rate increases exponentially with decreasing interaction energy, it is crucial to understand the response of nuclear recoils (NRs) in LXe down to the few keV energy scale. 
The XENON1T experiment, with a total mass of 3,200 kg, is currently the most sensitive dark matter search experiment in operation \cite{aprile2017first,aprile2017xenon1t}.
XENON1T and future detectors are pushing the sensitivity of the dark matter search to an unprecedented level, demanding high-precision understanding of the low-energy NR response.

Measurements of NR response in LXe have been carried out mainly using two methods: neutron fixed-angle scatters and data-simulation comparison of the NR spectrum, which are referred to as direct and indirect measurements, respectively, in this study.
Direct measurements \cite{aprile2005scintillation,prl2006,sorensen2009scintillation,aprile2009,manzur2010scintillation,horn2011nuclear,plante2011new} use one primary LXe detector surrounded by single or multiple neutron detectors to deduce the scattering angle from a known-energy source to reconstruct the deposited energy. Selecting this angle provides samples of pseudo-monoenergetic NRs.
Indirect measurements \cite[e.g.][]{xe100_nr} obtain the NR response by comparing a continuous data spectrum to the expected spectrum from simulations that include models of neutron propagation and detector response. 
These measurements thus rely more on the accuracy and precision of the models compared to direct measurements, which are more affected by detector systematic uncertainties.
Direct measurements are difficult to perform in large detectors built for dark matter detection since neutrons rarely escape the detector after interaction.
The only exception is LUX \cite{lux_nr}, which used a collimated neutron source and reconstruction of the deposited energy via multiple scatters.
Therefore, the interpretation of dark matter searches in large detectors relies on the global understanding of all direct measurements \cite{nest_nr}, and the validation of this understanding with in-situ calibration data \cite[e.g.][]{xe1t_sr1}.

For NRs with a few keV energy, there are three major channels for energy deposition: the excitation of xenon atoms, the formation of electron-ion pairs, and atomic motion \cite{lindhard1963integral}.
Excimers, dimers formed with an excited xenon atom (exciton), decay and produce photons.
Electrons and ions can either recombine, leading to the formation of an excimer, or remain separated.  
Atomic motion is not detected in LXe TPCs.
The relation between these channels is shown in \eqnref{eqn:lxe_energy_deposition}:
\begin{equation}
	\label{eqn:lxe_energy_deposition}
    \frac{(1-f_{\textrm{am}})E}{W} = N_{\textrm{ex}} + N_{\textrm{i}} = N_{\gamma}/f_l + N_{\textrm{e}} = N_q,
\end{equation}
where $E$ is the energy deposited by a NR, $W$ is the average energy required to excite or ionize a xenon atom in LXe, $f_{am}$ is the fraction of energy lost to atomic motion, and $f_l$ is the biexcitonic quenching (Penning quenching) factor.
$N_{\textrm{ex}}$, $N_{\textrm{i}}$, $N_{\gamma}$, $N_{\textrm{e}}$, and $N_q$ are the number of excimers, electron-ion pairs, scintillation photons, separated electrons, and total quanta, respectively.

Most of the previous direct and indirect measurements focus on studying either the light yield $L_y(E) = N_{\gamma}/E$ or charge yield $Q_y(E) = N_{\textrm{e}}/E$.
In \citeref{prl2006, manzur2010scintillation}, both $L_y$ and $Q_y$ are measured but their correlation is not exploited.
The correlation between light and charge signals is important for understanding the asymmetric stochastic processes at low energy.
Additionally, only \citeref{prl2006, manzur2010scintillation} have measured the effect of an electric field on the light and charge yields.
Unlike in the case of electronic recoils (ERs)~\citeref{goetzke2016measurement}, these measurements show little to no variation of the light and charge yield of NRs in LXe with an applied electric field.
However, additional measurements are necessary since recombination for low energy NRs is expected to be energy dependent as well as field dependent \cite{ti_recombination}, and both measurements cannot make conclusive statements below $\sim$45~keV.
In this work, we present the results of a new simultaneous measurement of the NR light and charge yields in LXe, using fixed-angle neutron scatters.
The data used in this work were taken at three electric fields, 0.19, 0.48, and 1.02\,kV/cm, with the nuclear recoil energy ranging from 3\,keV to 74\,keV.
Monte Carlo (MC) simulations, including models of the stochastic processes of light and charge production in LXe, as well as the detection and trigger efficiencies and detector response fluctuations, are used for parameter estimation.

\section{Experimental Setup and Operation}
\begin{figure}[htp]
    \centering
    \includegraphics[width=\columnwidth]{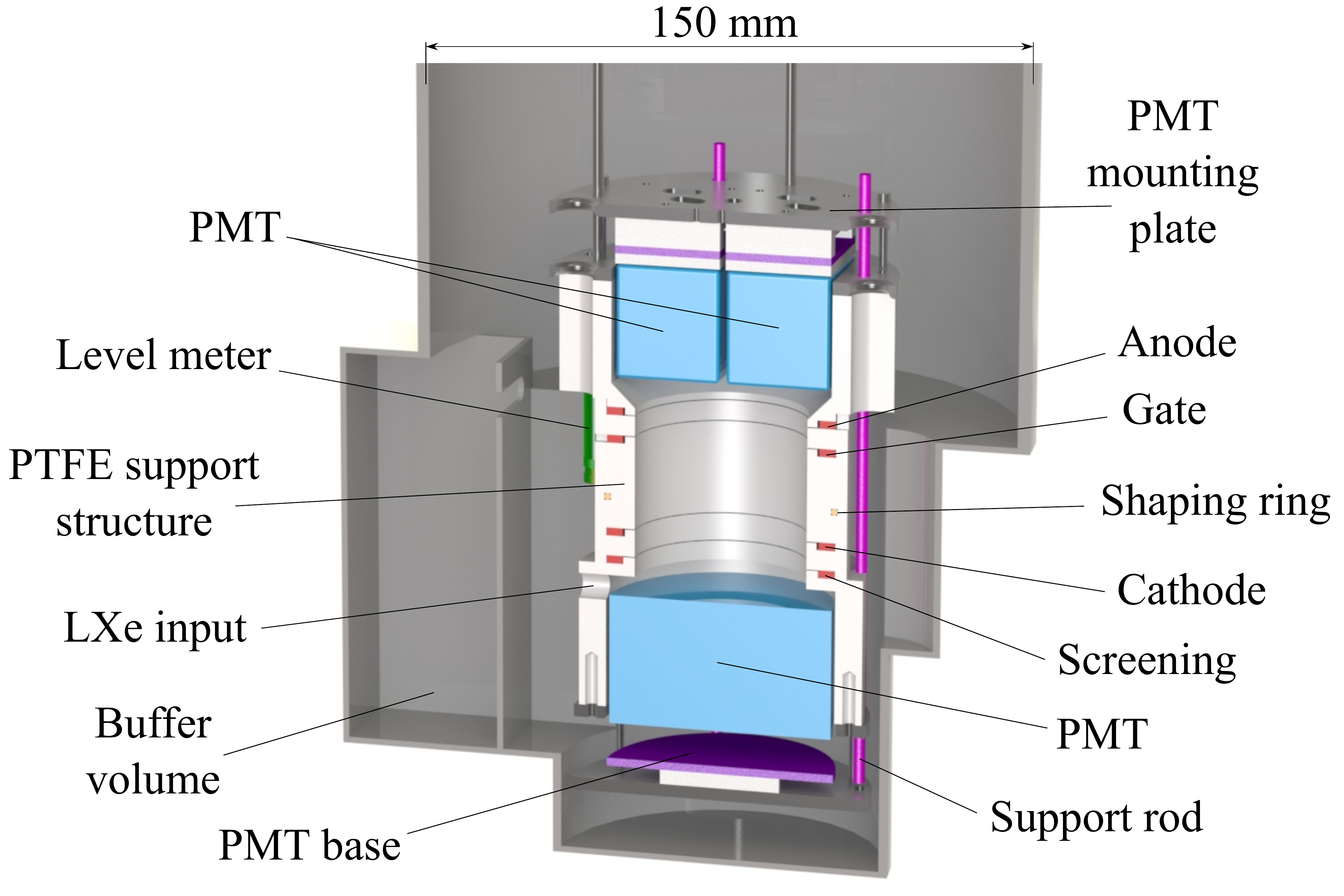}
    \caption{
    Schematic drawing of neriX TPC. 
    Plot from~\cite{goetzke2016measurement}.
    }
    \label{fig:nerix_tpc}
\end{figure}

The neriX (nuclear and electronic recoils in xenon) detector~\cite{goetzke2016measurement} was used for this study, which is a dual-phase, liquid and gaseous, xenon TPC. 
The TPC technique allows the simultaneous detection of the scintillation and ionization signals of an energy deposition.
The scintillation signal is directly collected by the photosensors placed in the detector, referred to as S1.
The ionization signal is amplified through electroluminescence of the drifted electrons in the gaseous xenon between the liquid-gas interface and anode.
The photons generated by the electroluminescence are then detected by the photosensors, referred to as S2.
More details of dual-phase TPC principle can be found in~\cite{aprile2010liquid}.
The schematic drawing of the neriX TPC is shown in Fig.~\ref{fig:nerix_tpc}.
Mesh electrodes are used for cathode, gate, and anode.
The TPC has a cylindrical sensitive (active) volume that is 23.4~mm in height (between the cathode and gate meshes, providing a drift field for electrons) and is 43~mm in diameter.
The distance between the gate to anode mesh is 5\,mm.
During operation, the liquid level is adjusted to be 2.5\,mm above the gate.
With +4.5\,kV applied to the anode, the electric field strength in the gaseous xenon is approximately 10\,kV/cm~\cite{luke_thesis}, based on COMSOL simulations~\cite{COMSOL}, for extracting electrons from the liquid and produce proportional scintillation.
The detector was designed to minimize the amount of inactive xenon and additional materials surrounding the active volume to minimize undetectable energy depositions that change the energy and direction of the incoming neutron and introduce systematic uncertainties during calibrations, making neriX optimal for measuring the signal response of LXe.
Four multi-anode Hamamatsu R8520-M4 photomultiplier tubes (PMTs) are placed on the top array of the TPC, for reconstructing the positions of events.
One 2$^{\prime\prime}$ Hamamatsu R6501 PMT is deployed at the bottom as the main light collector for energy reconstruction.
The cathode voltage can be tuned to produce electric fields between 0.15 kV/cm and 2.5 kV/cm, making neriX ideal for systematically measuring the field dependence of the light and charge yields for ERs~\cite{goetzke2016measurement} and NRs. 

During the fixed-angle measurements, the neriX detector was irradiated with 2.45\,MeV neutrons generated by a \ce{^{2}H}$(d,n)$\ce{^{3}He} generator~\cite{plante2011new}.
The neutron generator was placed 43\,cm away from the center of the neriX TPC.
Four 3$^{\prime\prime}$-diameter liquid scintillator (LS) detectors~\cite{ej301_manual} were placed at fixed positions around the detector.
The LS detectors have excellent gamma-neutron discrimination and provide high efficiency for tagging the outgoing neutrons from the neutron-Xe scatters inside the neriX TPC.
The LS detectors were positioned to a precision of $\sim$3\,mm in all three spatial dimensions using an auto-levelling laser device. 
The distances of LS detectors to the center of neriX TPC are listed in Table.~\ref{tab:nerix_ej_positions}.

The data acquisition (DAQ) system used in this measurement is similar to the one described in~\citeref{goetzke2016measurement}.
To trigger on S2 signals, which usually have a spread in time of about 1\,$\mu s$, the readout is triggered when an S2 has a time-over-threshold window larger than 0.4\,$\mu s$.
To avoid spurious S2 triggers, a 100\,$\mu s$ holdoff is applied, giving a 100\,$\mu s$ dead time to each valid S2 trigger.
For the measurement of fixed-angle neutron-Xe scatters, a coincidence between the TPC and one of the LS detectors signals within 29.5\,$\mu s$, which corresponds to the maximum electron drift time in the TPC, is required.
The signals from the PMTs are digitized by CAEN V1724 digitizers with a sampling frequency of 100\,MHz.

\begin{table}[h]
\footnotesize
\centering
\def\arraystretch{1.3}
\begin{tabular}{c||cccc}
\hline\hline
$\theta$ & $30^{\circ}$ & $35^{\circ}$ & $45^{\circ}$ & $53^{\circ}$ \\
Energy[keV] & $4.95 \pm 0.83$ & $6.60 \pm 1.52$ & $10.62 \pm 1.54$ & $13.95 \pm 2.46$ \\
LS distance [cm] & 76.7 & 42.0 & 59.0 & 37.0 \\
\hline\hline
\end{tabular}
\caption{The energies of the peaks and corresponding standard deviations which are derived via a detailed MC produced in Geant4 \cite{geant4}.
The distances of LS detectors with respect to the TPC center is also listed.
}
\label{tab:nerix_ej_positions}
\end{table}

\begin{figure}[b]
\centering
\includegraphics[width=\columnwidth]{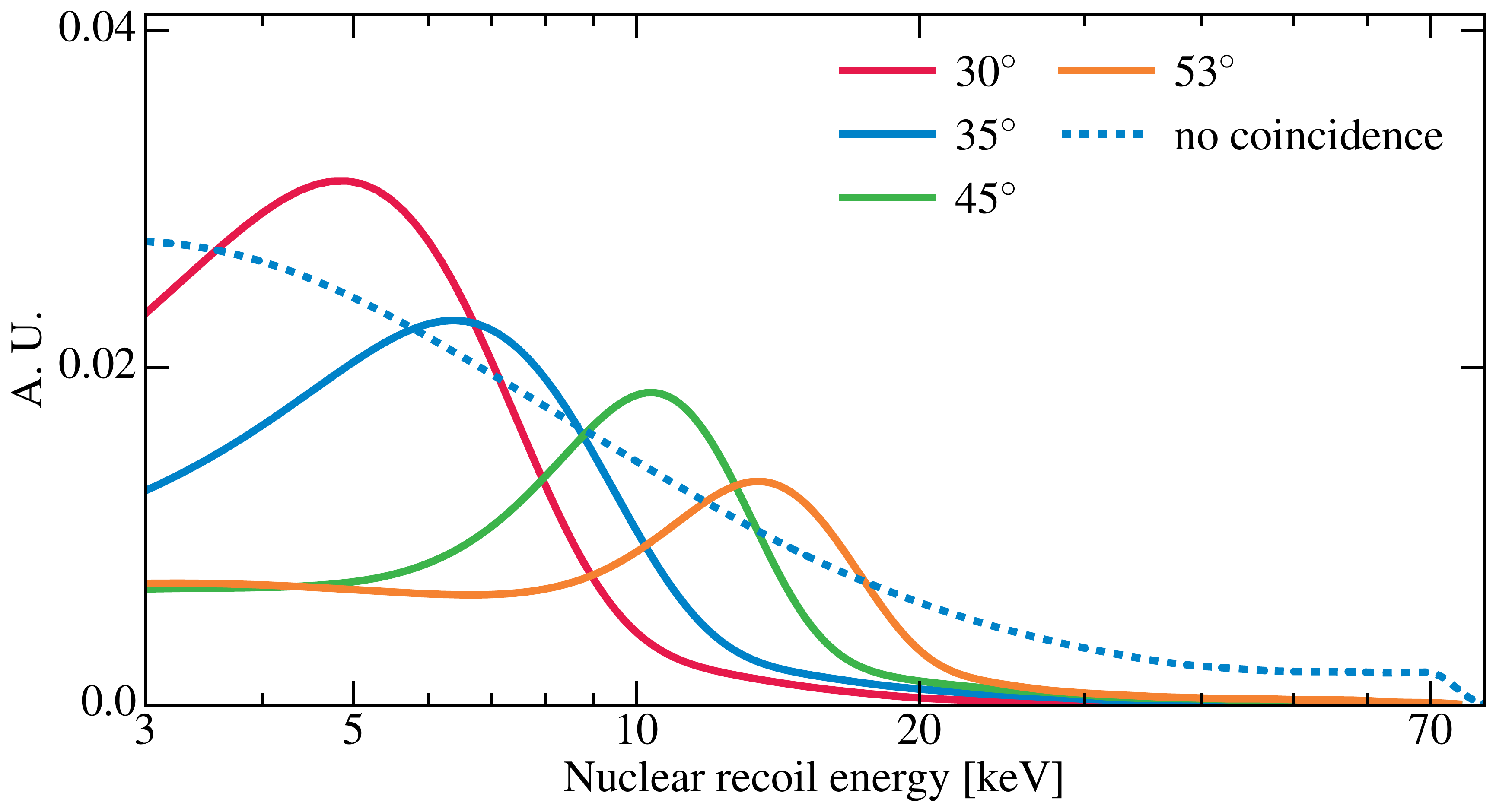}
\caption{
The deposit energy spectra of the neutron scatters with scattering angles of 30$^{\circ}$ (red), 35$^{\circ}$ (blue), 45$^{\circ}$ (green), and 53$^{\circ}$ (orange) in detector fiducial volume. 
The spectra are derived using Geant4 simulation toolkit~\cite{geant4}.
The nuclear recoil energy spectrum of neutron scatters with the coincidence requirement between TPC and LS detector trigger inhibited is also shown in dashed blue line.
}
\label{fig:energy_spectra}
\end{figure}

The fixed-angle neutron-Xe scatter data, later referred to as coincidence data, were collected at scattering angles of 30$^{\circ}$, 35$^{\circ}$, 45$^{\circ}$, and 53$^{\circ}$, corresponding to NR energies of approximately 5, 7, 10, and 14\,keV, respectively.
All coincidence data were taken at electric fields of 0.19, 0.48, and 1.02\,kV/cm.
The field values were derived using a COMSOL simulation~\cite{COMSOL}, similar to the procedure in~\citeref{goetzke2016measurement}.
During the entire data collection time, the PMT gains were stable within about a 2$\%$ level, monitored by bi-weekly LED calibrations.
Bi-weekly $^{22}$Na and $^{137}$Cs calibrations at different fields, including the scanned three fields, were performed for calibrating and monitoring the stability of the photon detection efficiency g$_1$ and charge amplification factor g$_2$ of the TPC.
We refer to g$_1$ (g$_2$) as the ratio of the number of reconstructed photoelectrons to the number of photons (electrons) in the scintillation (ionization) signal.
In addition to the coincidence data, data of neutron-Xe scatters without the coincidence requirement were taken, herein referred to as band data.
These data were taken to validate and tune the detector response model that will be introduced in later sections.
Fig.~\ref{fig:energy_spectra} shows the deposited energy spectra derived from Geant4~\cite{geant4} simulations, including the geometry for all detectors, for each fixed-angle as well as band data. The means and standard deviations of the energy depositions are summarized in Table~\ref{tab:nerix_ej_positions}.
The spread of the energy spectrum is caused by the finite sizes of the TPC and LS detectors, and by neutron scattering off detector materials before or after depositing energies in the active volume of the TPC.

\section{Data Analysis}

The transverse position of a scatter in the TPC is reconstructed with a neural network trained on optical simulations of the S2 pattern on the top PMTs (16 channels in total), which is a standard technique that is used in dual-phase TPCs~\cite{aprile2014analysis}.
The average X-Y position reconstruction resolution is about 0.9\,mm.
The Z position of the scatter is reconstructed as $Z = (T-T_g)/\nu_d$, where $T$ is the drift time of the scatter event, $T_g$ is the drift time of events that happen at the gate mesh, and $\nu_d$ is the electron drift velocity in LXe.
The drift time of events that happen at the gate mesh $T_g$ and at the cathode mesh $T_c$ are calibrated using S2s that are caused by the impurities, which are attached to the electrode and photoionized by S1 light, giving an electron drift velocity of $\nu_d = 23.4\,mm/(T_c-T_g)$.
The electron drift velocities at drift fields of 0.19, 0.49, and 1.02\,kV/cm are estimated to be 1.56, 1.77, and 2.00\,mm/$\mu$s, respectively, in our measurement.
To avoid potential non-uniformity of the electric field at the wall of the TPC, and near the gate and cathode meshes, events with reconstructed radius larger than 18\,mm and with reconstructed depth not in the range of 4 to 22\,mm below the gate mesh (the selected volume is refered to as the fiducial volume) were excluded in the analysis.
The data quality is further improved by removing events with spurious fraction of signal size on the top PMT array for both S1 or S2.
A time-of-flight requirement based on the reconstructed time of the TPC S1 signal and the LS signal is applied to reject accidental pileup of events.
In addition to the selection criteria above, pulse shape discrimination (PSD) of the LS detector is used to reject gamma background events.
A selection on the pulse height of the LS signal is used as well, to choose the LS signal region with high gamma rejection (larger than 99\%).


Due to the compact size of the LXe detector, the spatial dependencies of the S1 and S2 signals are not as significant as in large TPCs such as that of XENON1T~\cite{xe1t_instr}.
For S1s, the dependence arises from the fact that photons emitted from the top region of the TPC travel longer on average compared to those emitted from the bottom region because of total internal reflection of light at the liquid-gas interface, and thus have more chance to get absorbed by impurities prior to reaching the bottom PMT.
The standard deviation of the size of S1 signals in the fiducial volume is on the order of 10-20\%.
For S2s, it is due to the attachment of drifting electrons to electronegative impurities in the LXe, and is on order of 5$\%$.
The spatial dependencies of the S1 and S2 signals are monitored using $^{137}$Cs calibration data, and are found to be stable during the entire data taking period.
We correct the S1 and S2 signals for their Z dependence in the following analysis. 
The field dependence of electron attachment is taken into account in the S2 correction.

The anti-correlation between the light and charge signals in LXe~\cite{doke_ac} (the conservation of the total quanta produced at an interaction site in LXe before detection) allows for the calibration of g$_1$ and g$_2$ using monoenergetic gamma lines from $^{137}$Cs and $^{22}$Na calibration data at different fields.
In this work, the energy reconstruction utilizes the corrected S2 from the bottom PMT array and the corrected S1.
No significant time dependence of g$_1$ and g$_2$ was observed, and the run-average g$_1$ is $0.125\pm0.003$ and g$_2$ (for S2 from bottom PMTs) is $20.7\pm0.3$.
The g$_1$ is compatible with the expectation of 0.10 to 0.15, estimated through an optical simulation of the detector, manufacturer-provided PMT quantum and collection efficiencies, and the double photoelectron emission probability from~\cite{dpe_2015}.
The single electron gain $G$ of the detector, defined as the S2 size of one extracted electron, is estimated using single electron train data after large S2s.
The resulting $G$ is $22.9\pm0.6$, which implies an extraction efficiency of $0.903\pm0.028$.

\begin{figure}[htpb]
\centering
\includegraphics[width=\columnwidth]{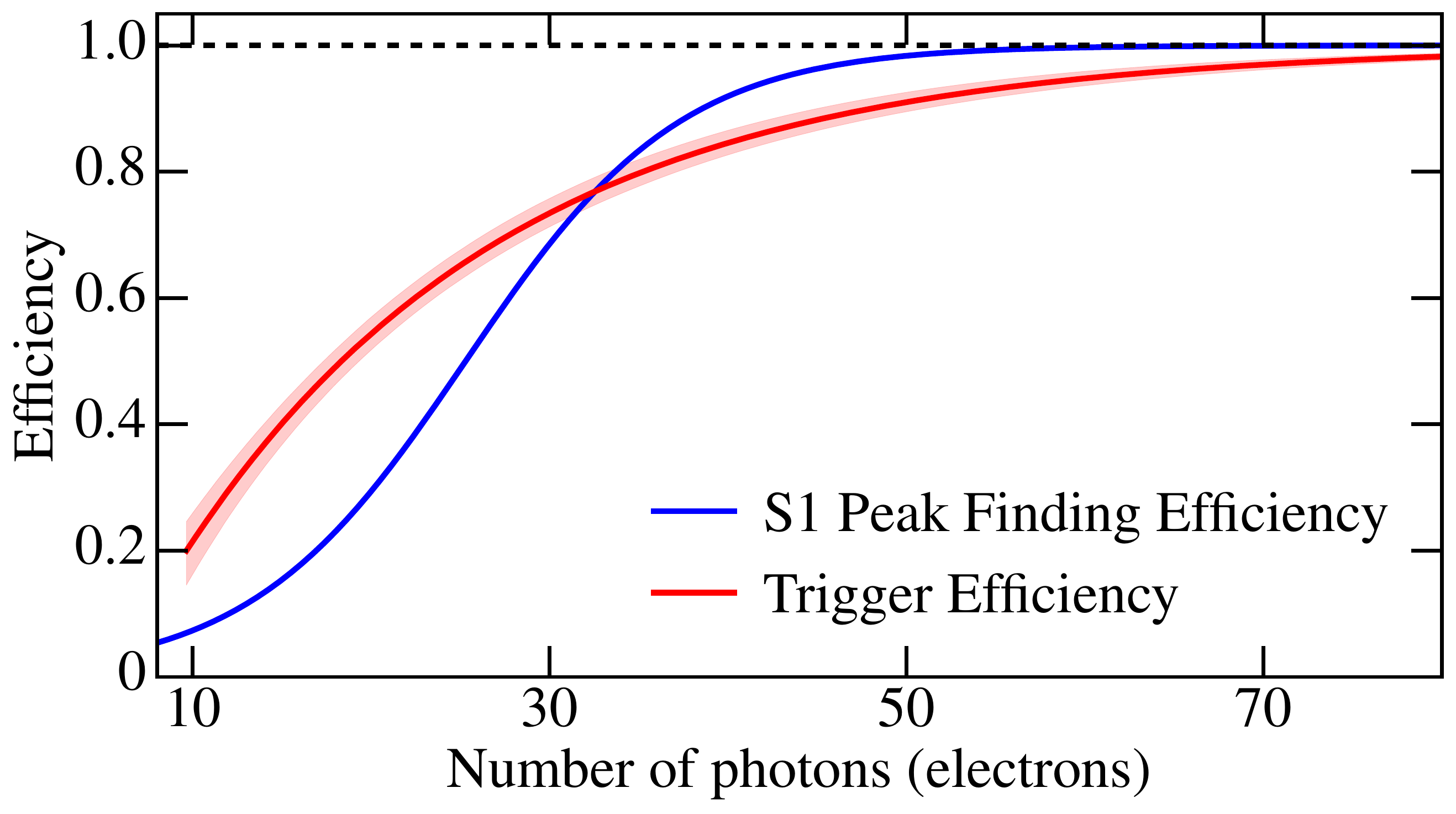}
\caption{
The trigger efficiency (red) as a function of number of electrons and the peak finding efficiency (blue) as a function of number of photons.
The solid lines are the median efficiencies.
The shaded regions show the uncertainty (15.4\% to 84.6\% credible region).
}
\label{fig:efficiencies}
\end{figure}

The main efficiency losses for the measurement come from the trigger efficiency of S2 and the peak finding efficiency of S1, and are considered in the analysis.
The trigger efficiency is related to the requirement of the S2 width in the DAQ setup.
During data taking, the trigger signal for each S2 peak is digitized.
With such digitized trigger signals, the trigger efficiency was evaluated using the S2s in randomly triggered data.
Peak finding efficiency loss is caused by a threshold in the S1 peak finding procedure, and is strongly related to the noise level of each PMT channel.
The peak finding efficiency is evaluated using a waveform simulation that includes the pulse shape of scintillation signals in LXe~\cite{s1_shape} and realistic noise samples.
The trigger and peak finding efficiencies are shown in Fig.~\ref{fig:efficiencies}.

\section{Method}

In previous fixed-angle neutron scatter measurements~\cite{aprile2005scintillation,prl2006,sorensen2009scintillation,aprile2009,manzur2010scintillation,horn2011nuclear,plante2011new}, a simple approximation of monoenergetic neutron scatters with some energy uncertainty was typically used.
Then a simple empirical signal smearing was applied to address the effect of statistical and systematic fluctuations.
As shown in Fig.~\ref{fig:energy_spectra}, there is some variance of the deposited energy of neutron scatters, presumably due to multiple scatters of neutrons off detector materials prior to reaching a LS detector, and due to the finite size of the active volume allowing a variation of scattering angles.
Thus, the assumption of a monoenergetic neutron scatter does not apply well to the analysis of neriX data where light and charge signals are simultaneously considered.
In addition, uncertainties in the light and charge measurements and in the scattering angles are all correlated, affecting the precision of energy reconstruction.
We perform a simultaneous fit, considering these correlations and energy variance, to all available neutron scatter data (both coincidence and band data) taken at different electric fields. The fitted model includes a parameterization of the energy dependence of the light and charge yields in LXe (microphysics described below).
During neutron scatter data taking, material activation by neutrons introduced gamma-induced ER backgrounds.
The ER background is negligible in coincidence data because of the coincidence requirement between the TPC and LS detectors, and because of the PSD selection of LS detector signals.
However, it is non-negligible in band data, for which we only select neutron scatter events with S1s larger than 40\,PE where there is good discrimination between ER and NR.
This high S1 threshold does not weaken the low energy analysis, since the inclusion of band data is mainly for constraining high-energy NRs, especially using the energy cutoff at $\sim$74\,keV, shown in Fig.~\ref{fig:energy_spectra}, arising from the backscatters of the 2.45\,MeV neutrons.
We will also present the results with mono-energy neutron scatter approximation (similar to~\cite{plante2011new}) for reference.

In the fit to data, we use the MC simulation-based detector response model for predicting the S1-S2 distribution of the neutron scatters.
The model includes the stochastic processes in the LXe microphysics of the production of light and charge, their propagation through the TPC, photon detection, and the efficiencies and biases of event reconstruction.
This approach is similar to that of~\citeref{xe100_tritium}, but with some simplification and extra terms to address the difference in detectors.

\begin{figure*}[!p]
	\centering

	\includegraphics[width=0.45\textwidth]{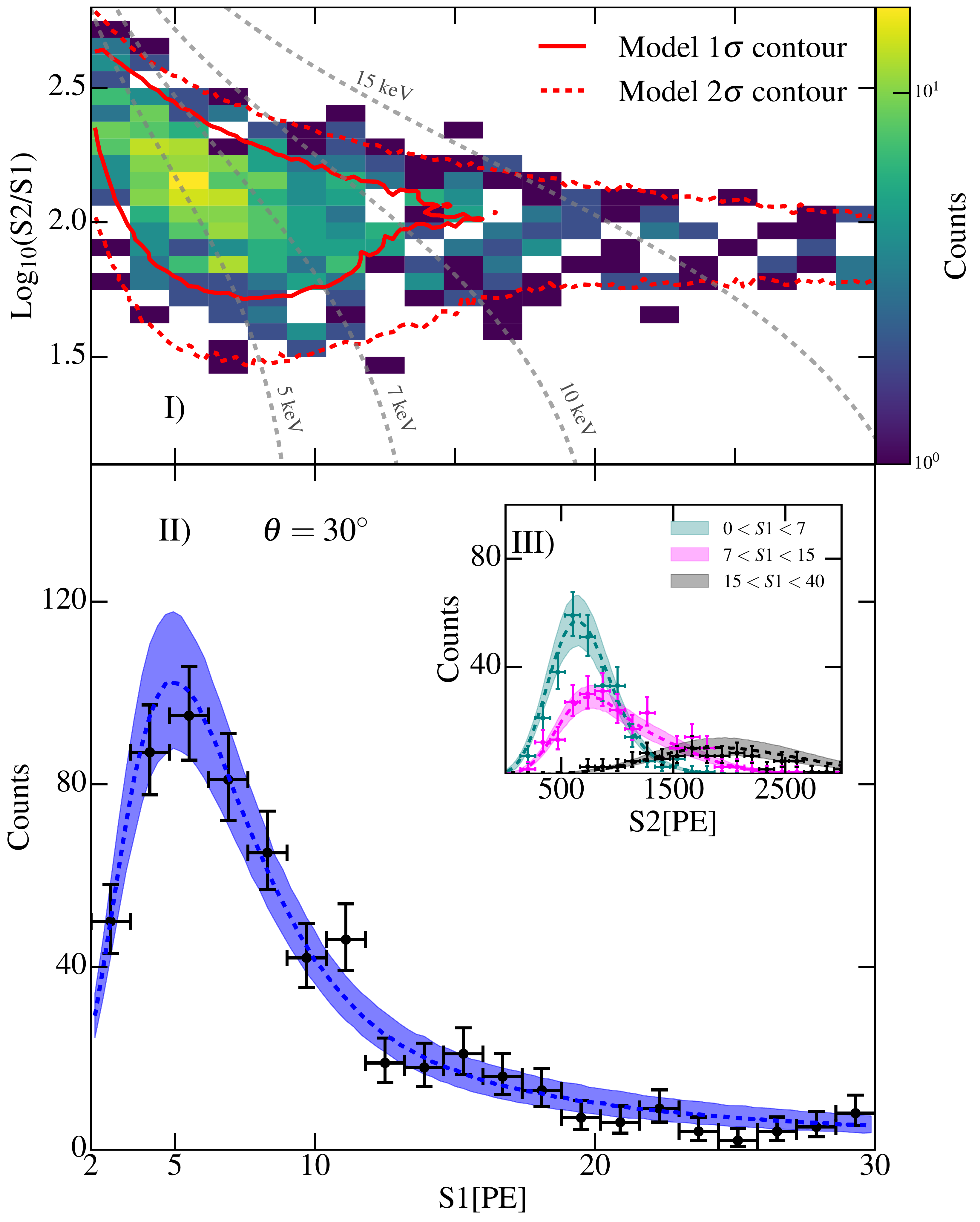}
	\includegraphics[width=0.45\textwidth]{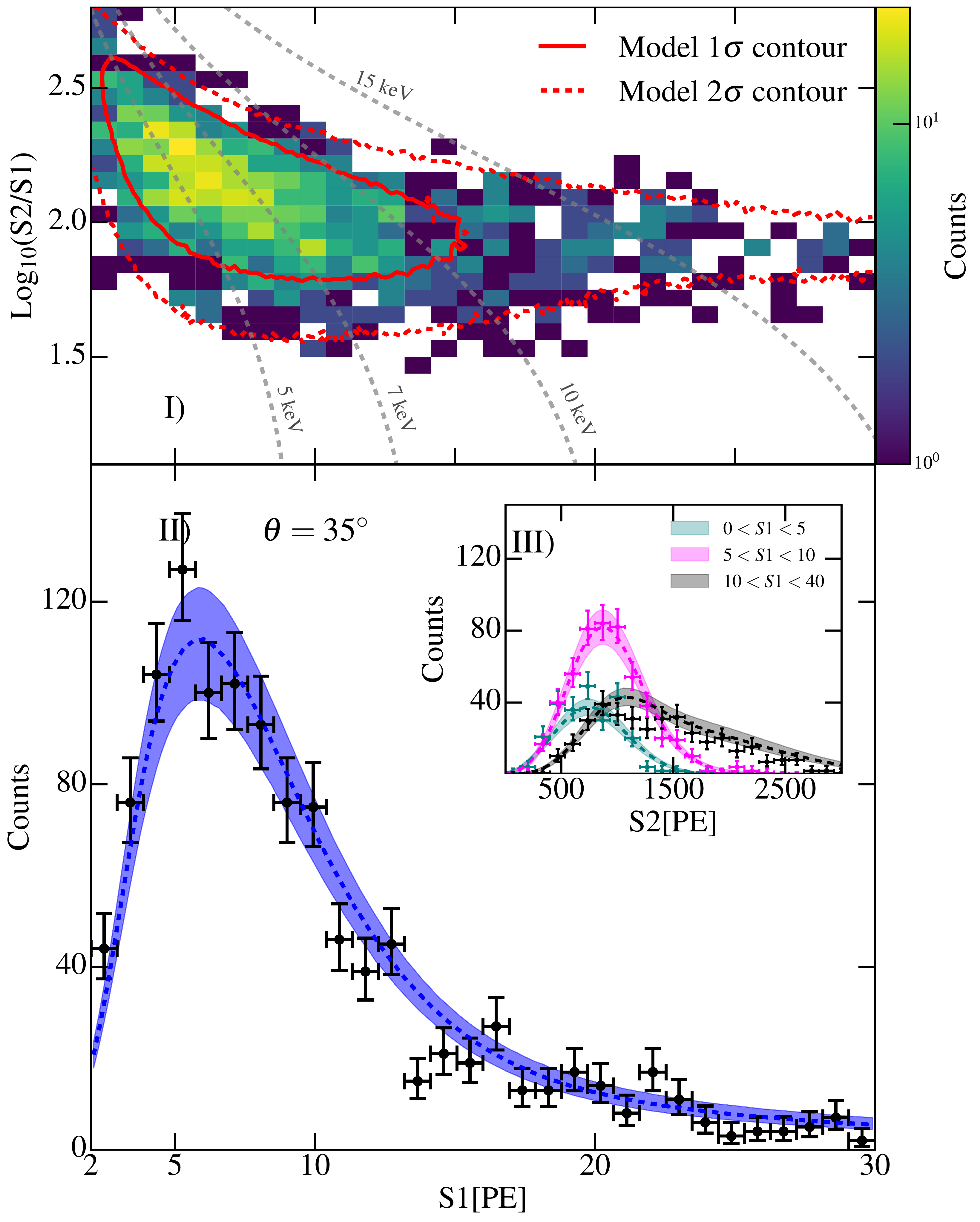}

	\centering
	
	\includegraphics[width=0.45\textwidth]{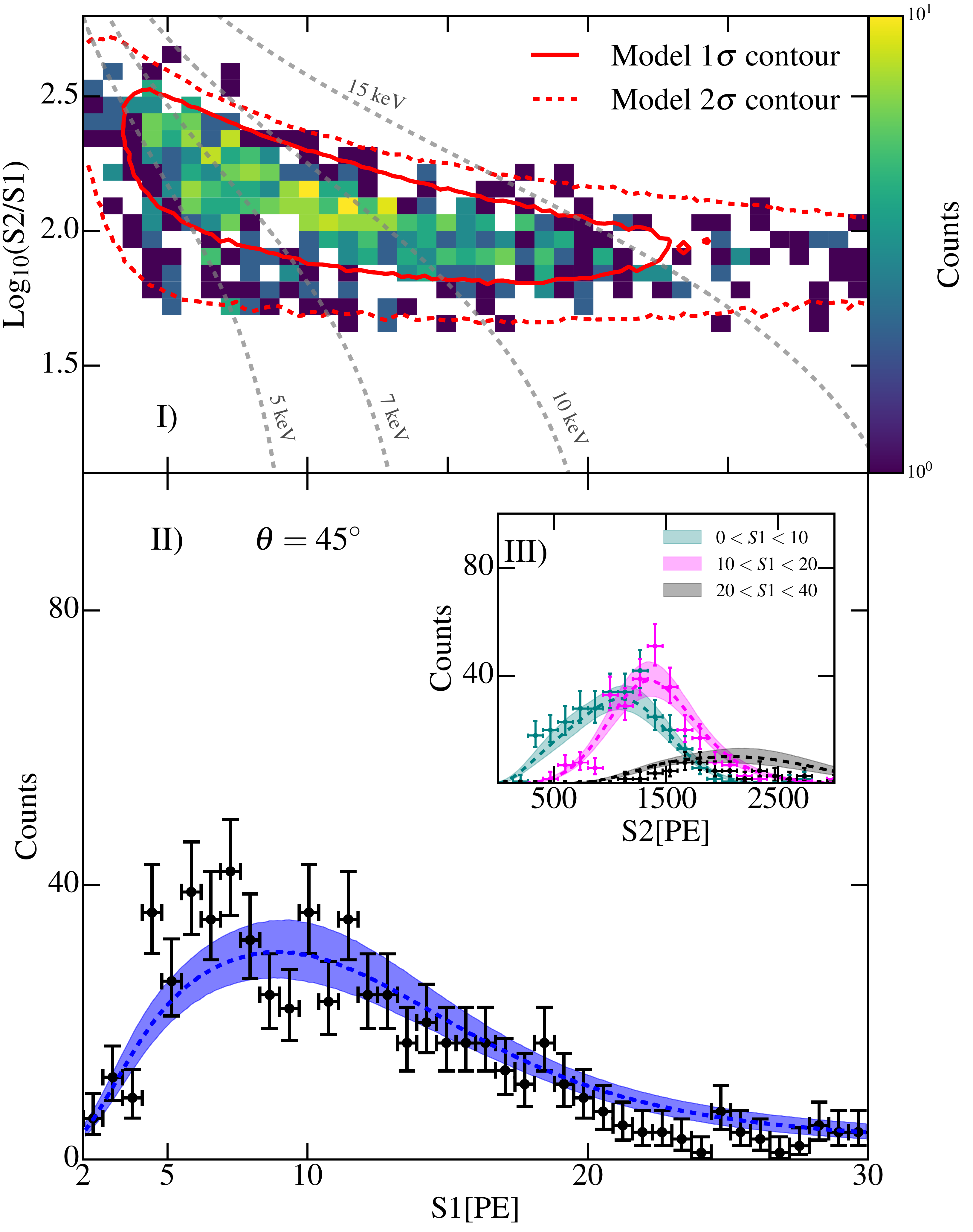}
	\includegraphics[width=0.45\textwidth]{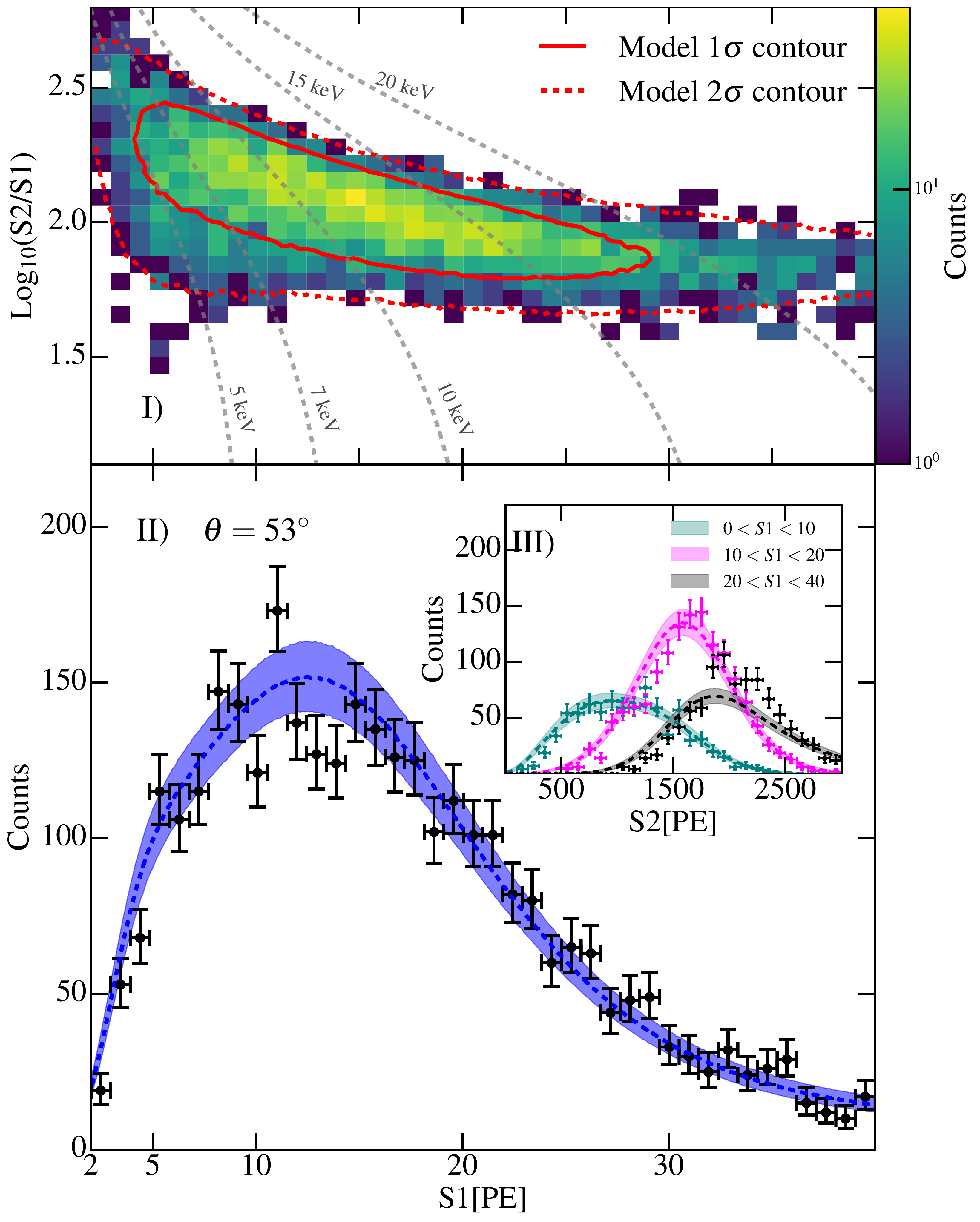}

	\centering
	\caption{
	Coincidence data taken with a drift field of 190 V/cm for scattering angles of 30$^{\circ}$, 35$^{\circ}$, 45$^{\circ}$, and 53$^{\circ}$, from left to right and top to bottom.
	Panel I) shows the data distribution on log$_{10}$(S2/S1) versus S1 space, overlaid with the 68\% (1$\sigma$, solid) and 95\% (2$\sigma$, dashed) red contour lines derived from the median of the 2D marginalized posterior from the MCMC sampling of the likelihood.
	Panel II) shows the S1 spectrum of data (black circles) compared to the 1D marginalized posterior (blue).
	The dashed line represents the median of the posterior, and the shaded region is the 15.4\% to 84.6\% credible region.
	Panel III) shows the S2 spectra in different S1 ranges. The data spectra are shown with the error bars, and the posterior medians and credible regions (15.4\% - 84.6\%) are shown in dashed lines and shaded regions, respectively.
	}
	\label{fig:nerix_nr_best_fit_1}
\end{figure*}

\begin{figure*}[!phtb]
	\centering

	\includegraphics[width=0.45\textwidth]{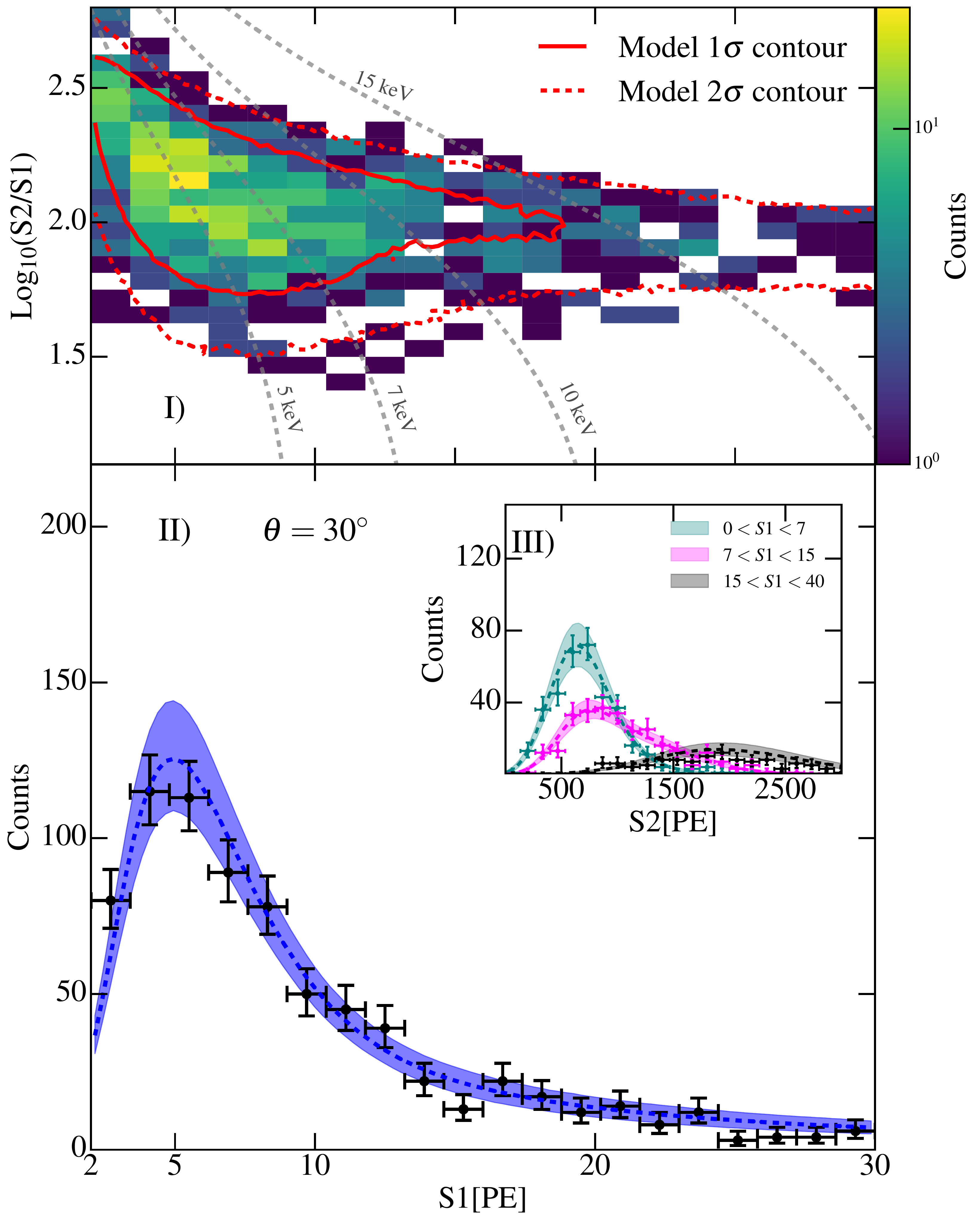}
	\includegraphics[width=0.45\textwidth]{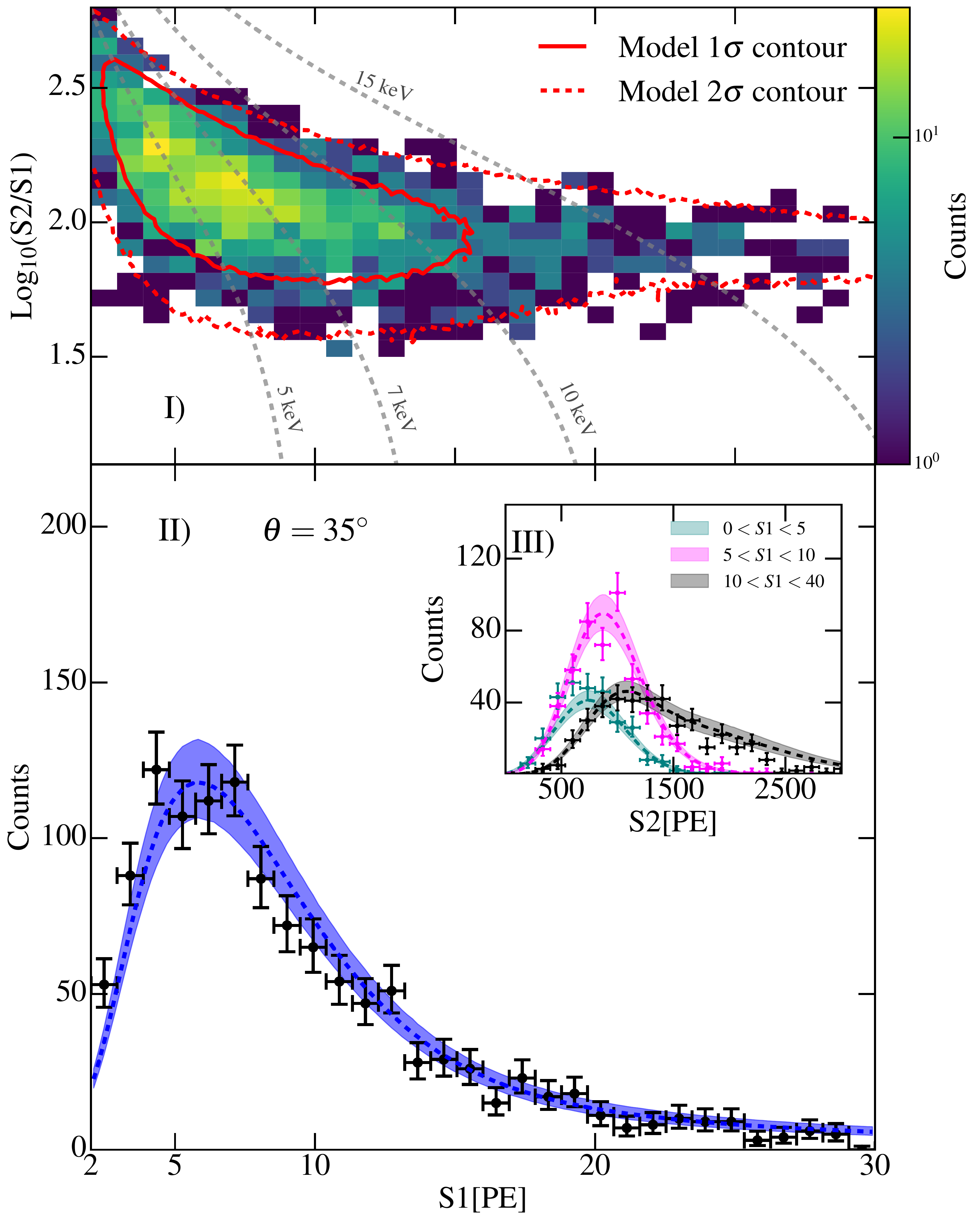}

	\centering
	
	\includegraphics[width=0.45\textwidth]{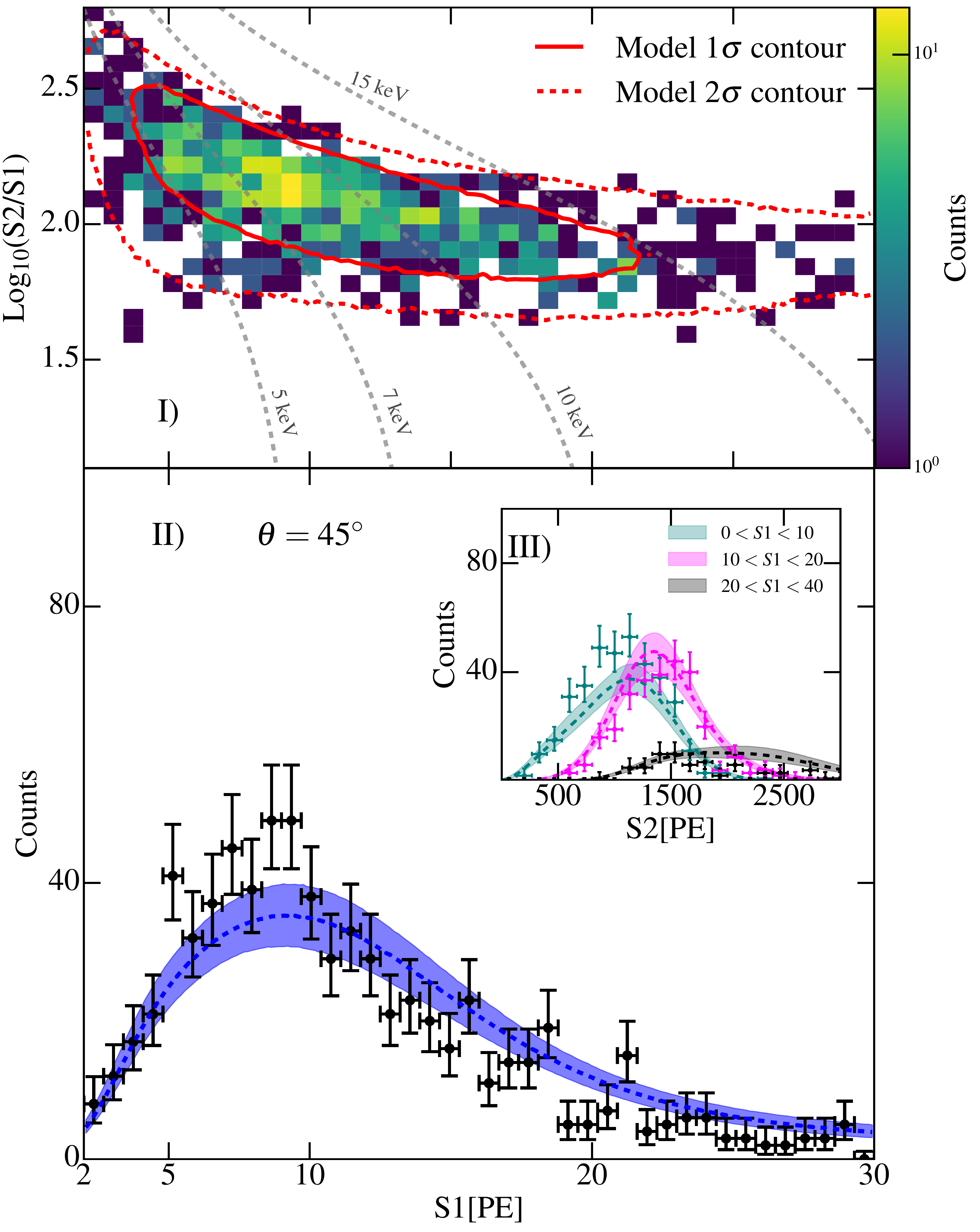}
	\includegraphics[width=0.45\textwidth]{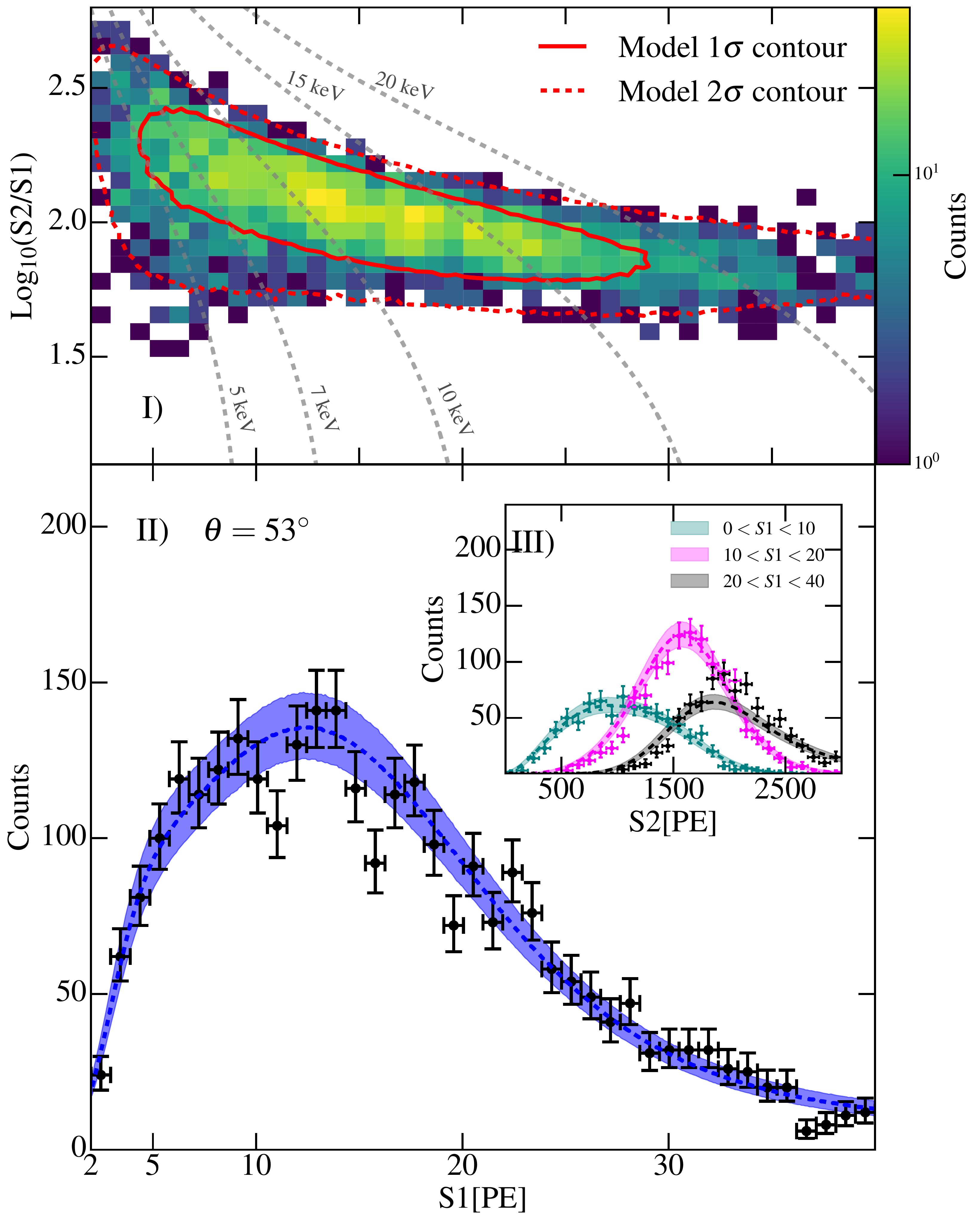}

	\centering
	\caption{
	Coincidence data taken with a drift field of 490 V/cm. The figure description is the same as in \figref{fig:nerix_nr_best_fit_1}.}
	\label{fig:nerix_nr_best_fit_2}
	\vspace{0.8in}
\end{figure*}

\begin{figure*}[!phtb]
	\centering

	\includegraphics[width=0.45\textwidth]{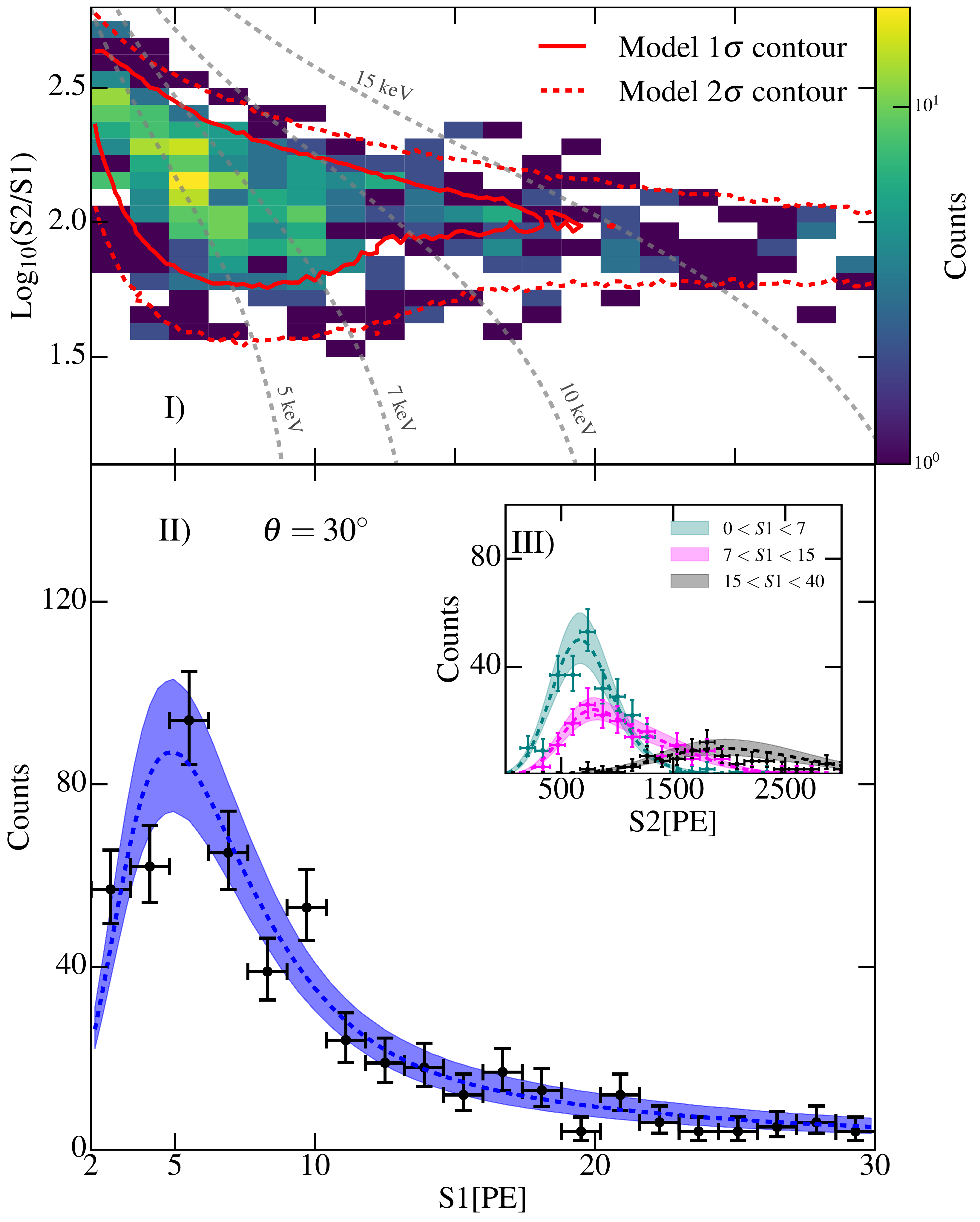}
	\includegraphics[width=0.45\textwidth]{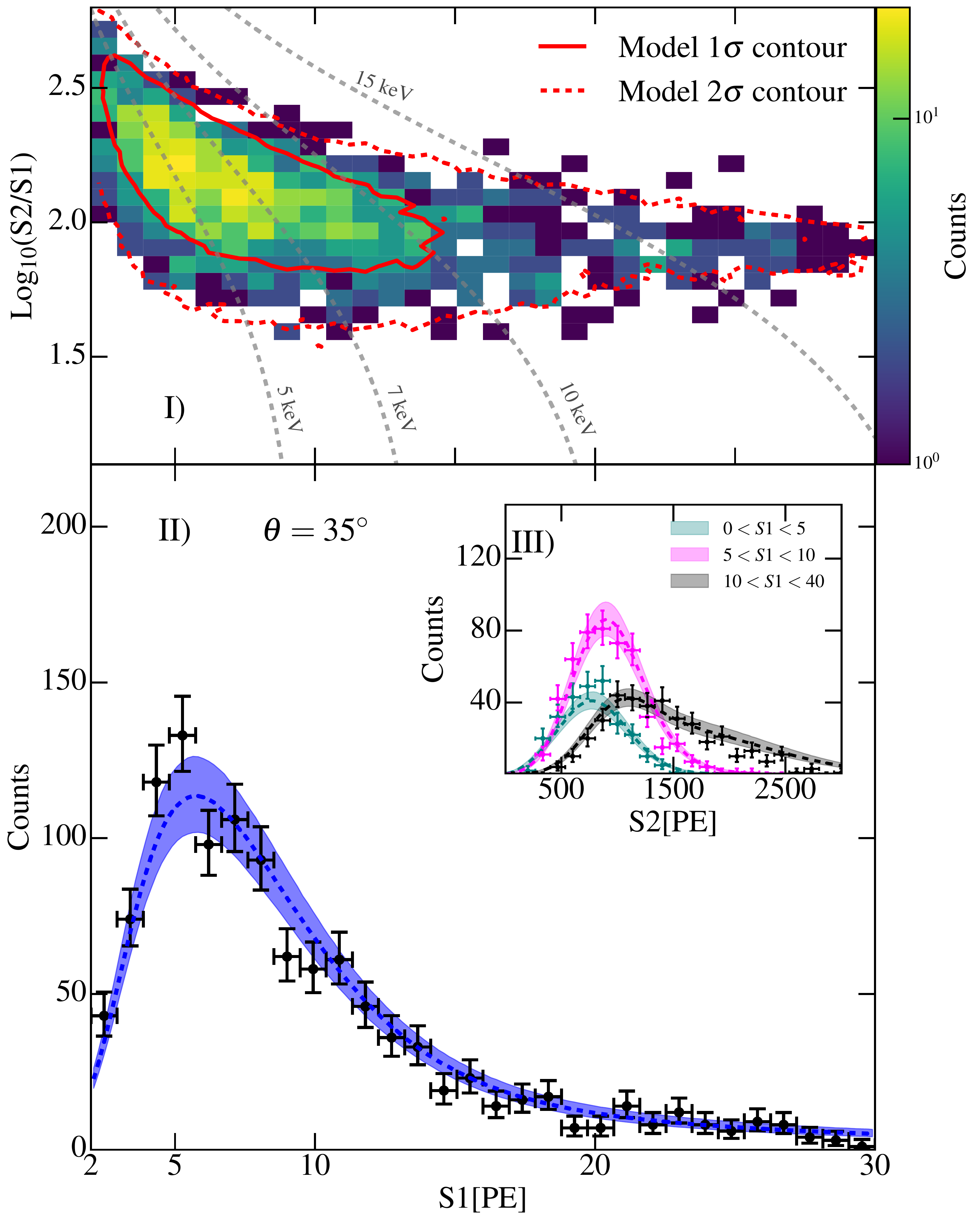}

	\centering
	
	\includegraphics[width=0.45\textwidth]{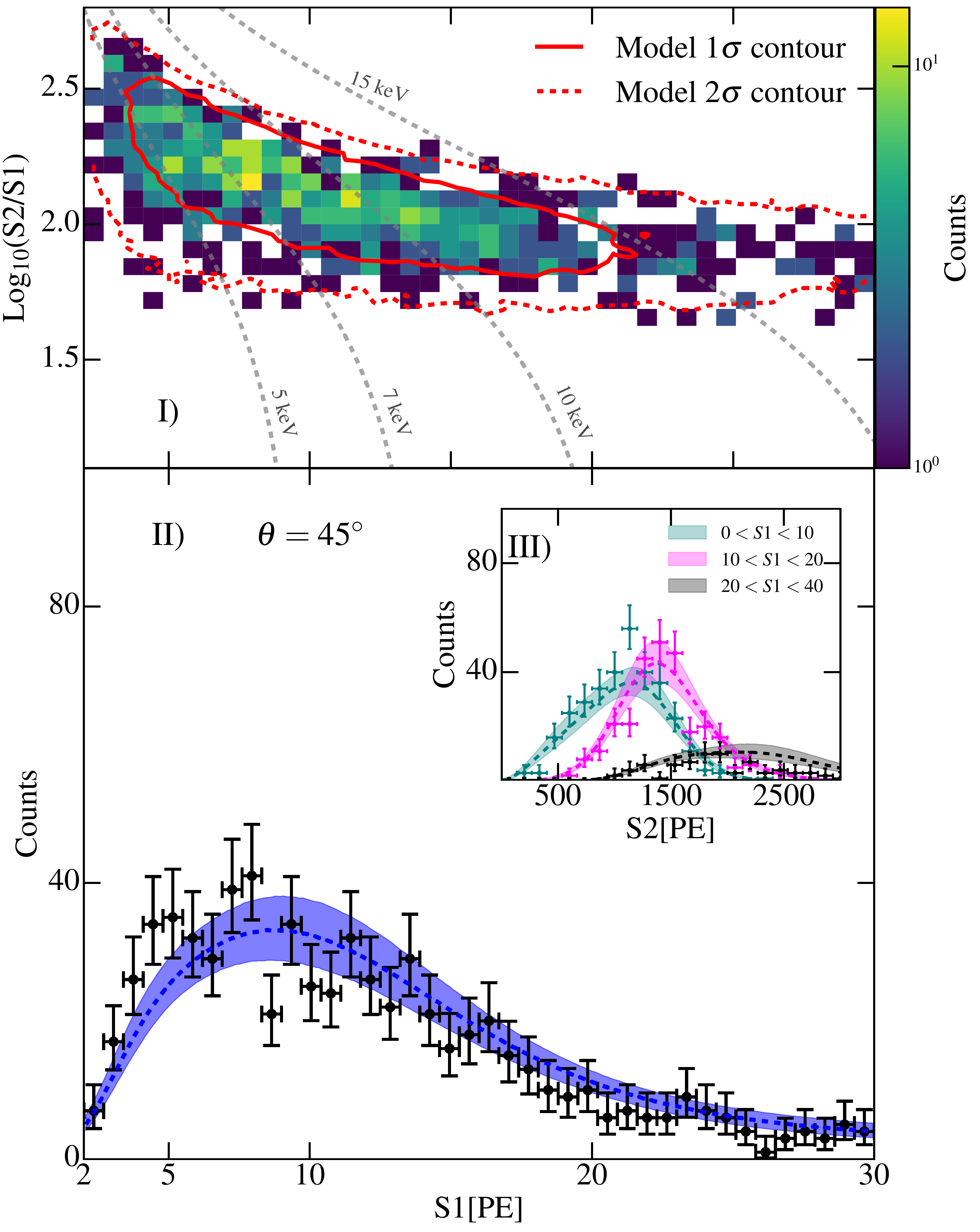}
	\includegraphics[width=0.45\textwidth]{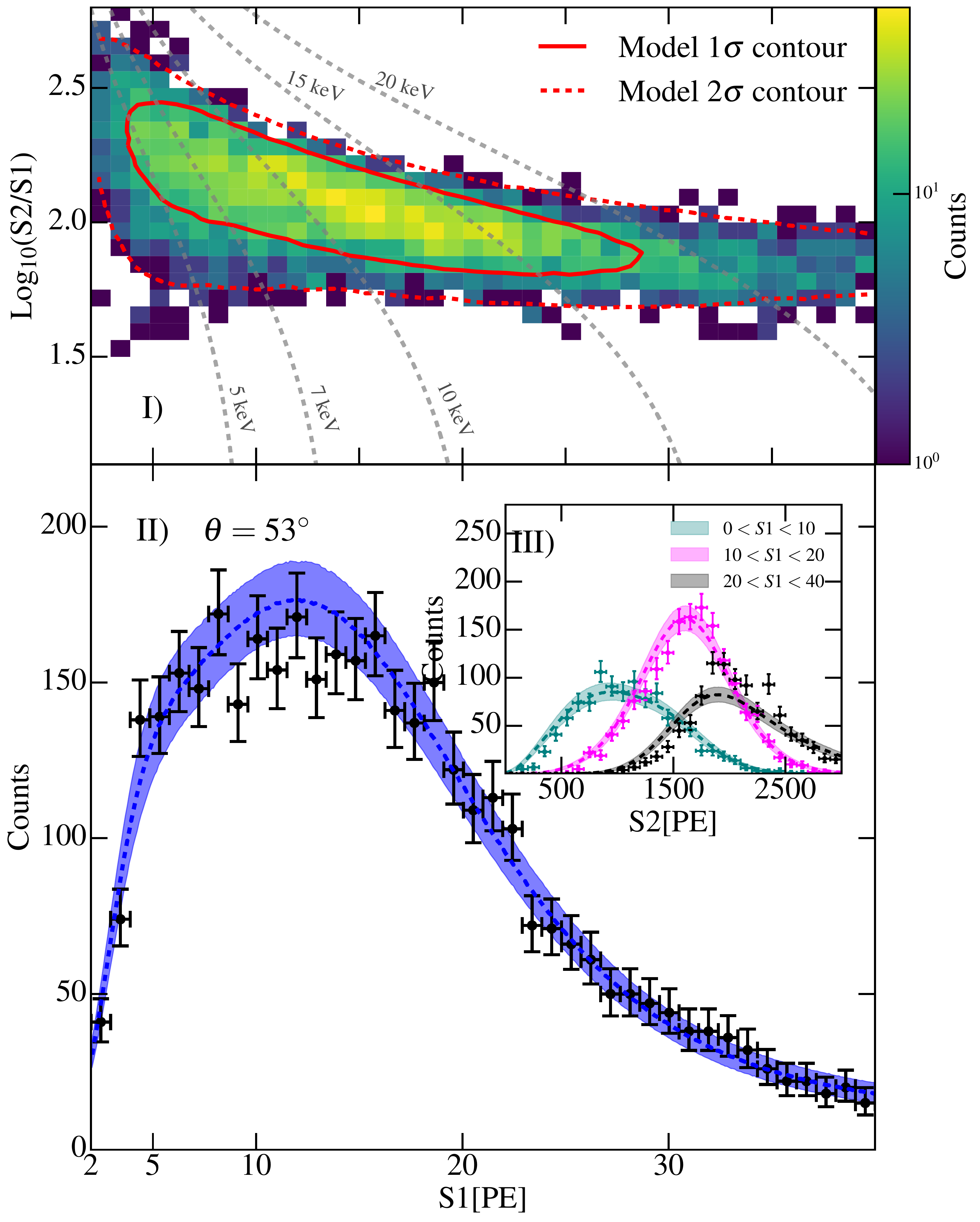}

	\centering
	\caption{
	Coincidence data taken with a drift field of 1.02\,kV/cm. The figure description is the same as in \figref{fig:nerix_nr_best_fit_1}.}
	\label{fig:nerix_nr_best_fit_3}
	\vspace{0.8in}
\end{figure*}

\begin{figure*}[htp]
	\centering

	\includegraphics[width=0.32\textwidth]{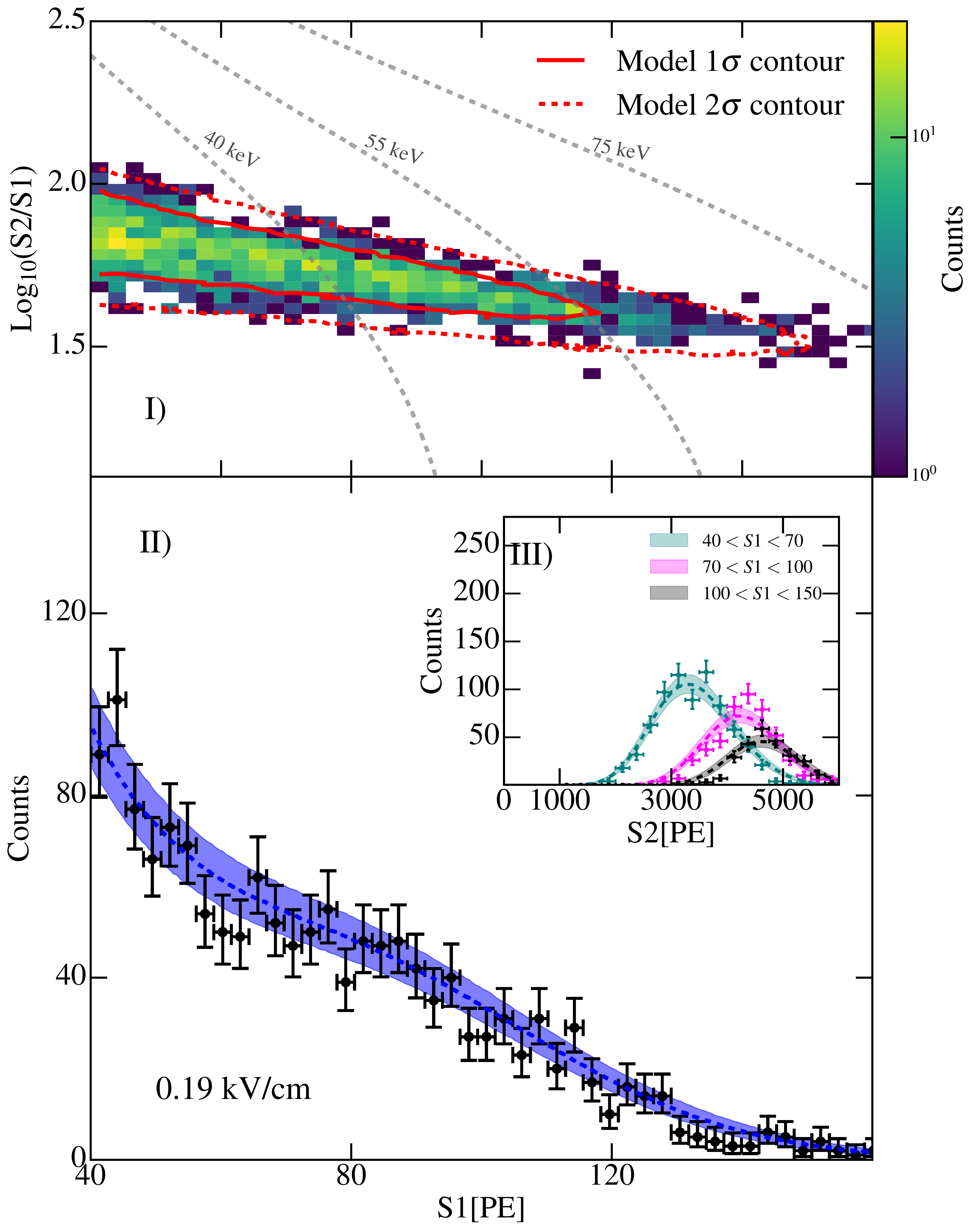}
	\includegraphics[width=0.32\textwidth]{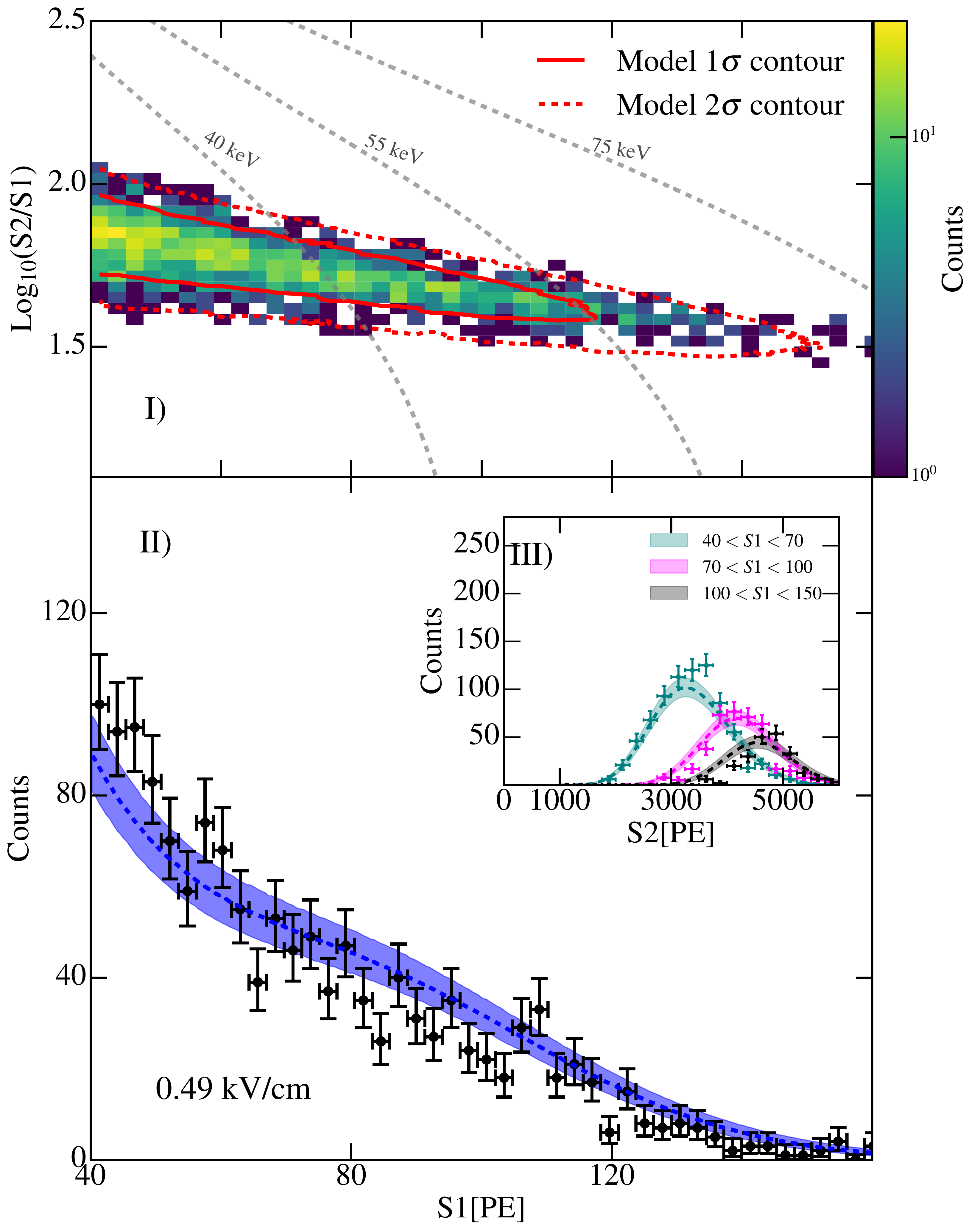}
	\includegraphics[width=0.32\textwidth]{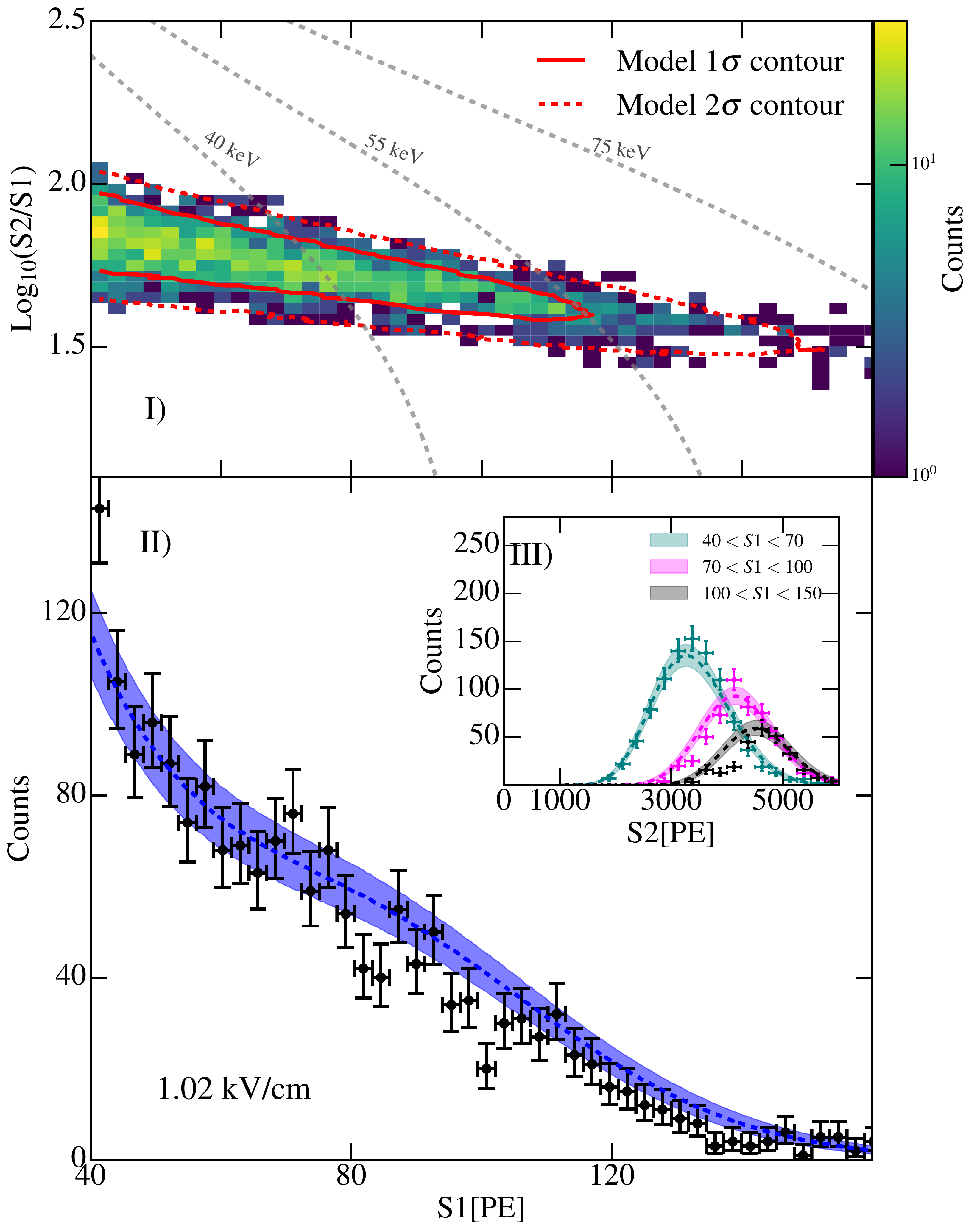}\\

	\centering
	\caption{
	Band data taken with scanned drift fields of 190\,V/cm, 490\,V/cm, and 1.02\,kV/cm, from left to right. The figure description is the same as in \figref{fig:nerix_nr_best_fit_1}.}
	\label{fig:nerix_nr_best_fit_4}
\end{figure*}

\subsection{Parameterization of the light and charge yields}
\label{subsec:lxe_microphysics}
The parameterizations of light and charge yields in LXe, which are the main outcomes of this experiment, are similar to that used in NEST v1.0~\cite{nest_nr} except for the removal of the field dependence since it will be probed in this measurement.
For each energy deposition, the energy is converted to the number of quanta $N_q$ (the sum of numbers of photons $N_\gamma$ and electrons $N_e$ produced, as in \eqnref{eqn:lxe_energy_deposition}) by $\langle N_q \rangle = \varepsilon \cdot L / W$, where $\varepsilon$ is the deposited energy, $L$ is the Lindhard factor~\cite{lindhard1963integral}, and $W$ is the average energy required to produce one photon or electron in LXe~\cite{nest_v0.98}.
The Lindhard factor is usually parameterized as:
\begin{equation}
\begin{array}{rl}
L(\epsilon) & = \frac{k g(\epsilon)}{1+k g(\epsilon)} \mathrm{,~where} \\
g(\epsilon) & = 3\epsilon^{0.15} + 0.7\epsilon^{0.6} + \epsilon,  \\
\epsilon & = 11.5 \left( \frac{\varepsilon}{keV} \right) Z^{-7/3},  \\
\end{array}
    \label{eq:Lindhard_Parameterization}
\end{equation}
$Z$ is the atomic number of the target material, and $k$ is a proportionality constant and the key model parameter.
Due to the nature of a recoiling Xe atom, the deposited energy is mostly converted into heat and the Lindhard factor is around 0.1 to 0.2.
The signal fluctuations arising from this heat loss is Poissonian and can be written as:
\begin{equation}
N_q \sim Pois(\langle N_q \rangle).
    \label{eq:heat_loss_fluc}
\end{equation}
A fraction of the quanta is in the form of xenon excimers that eventually de-excite and emit photons.
The rest is in the form of electron-ion pairs, part of which recombine to produce photons.
The fluctuation in the number of ions $N_i$ (and thus also the number of excimers $N_{ex}$, due to \eqnref{eqn:lxe_energy_deposition}) is binomial and can be written as:
\begin{equation}
N_i \sim Binom\left(N_q, \frac{1}{1+N_{ex}/N_i}\right).
    \label{eq:excimer-ion-ratio}
\end{equation}
The excimer-to-ion ratio $N_{ex}/N_i$ is parameterized similarly to~\citeref{nest_nr} as:
\begin{equation}
N_{ex}/N_i = \alpha (1 - e^{-\beta \varepsilon}),
    \label{eq:excimer-ion-ratio-parameterization}
\end{equation}
where $\beta$ is a constant and $\alpha$ depends on the electric field strength.
The recombination of electron-ion pairs induces additional fluctuation on $N_{\gamma}$ (and thus also $N_e$, due to \eqnref{eqn:lxe_energy_deposition}) through a binomial process:
\begin{equation}
N_e = Binom(N_q, 1-r),
    \label{eq:recomb_binom_process}
\end{equation}
where $r$ is the recombination fraction that can be both energy and field dependent. This fraction is assumed to follow the Thomas-Imel box model~\cite{ti_recombination} for low-energy NRs,
\begin{equation}
r = 1 - \frac{ln(1+N_i \varsigma)}{N_i \varsigma}, 
    \label{eq:recomb_frac}
\end{equation}
where $\varsigma$ is the Thomas-Imel constant and is allowed to be free for different electric fields.
For high-energy NRs, Penning quenching results from two excimers colliding into one, reducing the number of excimer de-excitations that would have otherwise produced light.
The Penning quenching $f_l$ can be parameterized as:
\begin{equation}
f_l = \frac{1}{1+\eta \varepsilon^{1/2}},
    \label{eq:penning_quenching}
\end{equation}
where $\eta$ is a free parameter and the index of 1/2 comes from the electron stopping power in LXe~\cite{bezrukov2011interplay}.
Consequently, an additional binomial fluctuation affects the number of de-exciting excimers, including the excimers formed in the ion-electron recombination process:
\begin{equation}
N_{\gamma} \sim Binom(N_{ex}+N_i - N_e, f_l).
    \label{Penning_fluc}
\end{equation}
The Penning quenching is assumed to be field-independent in this study.

\subsection{Light/Charge Detection and Reconstruction}
\label{subsec:detector_physics}
The fluctuation in the light collection process is modeled as:
\begin{equation}
\begin{array}{cc}
N_{hit} \sim & Binom(N_{\gamma}, g_1 / (1 + p_{dpe}) ), \\
N_{PE} \sim & N_{hit} + Binom(N_{hit}, p_{dpe}),
\end{array}
    \label{eq:light_collection_fluc}
\end{equation}
where $N_{hit}$ is the number of photons that reach the PMTs.
This differs from the number of photoelectrons (PE) $N_{PE}$ produced at the PMT photocathode due to the double-PE effect~\cite{dpe_2015}, where $p_{dpe}$ is the probability of double-PE emission.
The PEs are amplified in each PMT, digitized by the DAQ electronics, and reconstructed by software.
The amplification and electronics produce additional signal fluctuations:
\begin{equation}
\mathrm{S1} \sim Norm(N_{PE}, \sigma^2 = \sigma_{SPE}^2 N_{PE} + \sigma_{\mathrm{S1}}^2),
    \label{eq:s1_sys_fluc}
\end{equation}
where $\sigma_{SPE}$ is the intrinsic resolution of a PMT during the amplification of a single PE and is calibrated to be 0.645$\pm$0.006\,PE.
The fluctuations due to PMT amplification decrease as $\sqrt{N_{PE}}$.
Noise induced by the signal (signal-correlated) and electronic noise can further introduce fluctuations in the reconstructed energy.
Signal-correlated noise can include the PMT afterpulses and the photoionization of impurities.
We model the relative fluctuations from signal-correlated noise as a diminishing exponential of the signal size plus a constant:
\begin{equation}
\sigma_{\mathrm{S1}}/N_{PE} = a + b \cdot e^{-\mathrm{S1}/c},
    \label{eq:sys_fluc_parameterization}
\end{equation}
where $a$, $b$, and $c$ are free parameters.
The spatial dependence of S1 is neglected in the simulation since it is subdominant and the correction is applied to the data.
For charge collection, the statistical fluctuation from electron extraction at the liquid-gas interface is accounted for as:
\begin{equation}
N_{ext} \sim Binom(N_e, p_{ext}),
    \label{eq:extraction_fluc}
\end{equation}
where the fluctuation of the number of extracted electrons $N_{ext}$ is related to the extraction efficiency $p_{ext}$.
Electron loss due to impurity attachment is not considered in the simulation since it is negligible in such a compact detector as neriX.
The fluctuation of the S2 signal consists of two parts, one from single electron amplification in the gaseous phase, which decreases with $\sqrt{N_{ext}}$, and the other arising through software reconstruction from noise, spatial dependence of the signal, and the like,
\begin{equation}
\mathrm{S2} \sim Norm(G N_{ext}, \sigma^2 = \sigma^2_G N_{ext} + \sigma_{\mathrm{S2}}^2).
    \label{eq:s2_sys_fluc}
\end{equation}
$\sigma_G$ is the spread in single electron amplification gain $G$ and is estimated to be 8.59$\pm$0.32\,PE/e$^{-}$. 
The rest of the fluctuations in S2, $\sigma_{\mathrm{S2}}$/(GN$_{ext}$), are parameterized similarly to Eqn.~\ref{eq:sys_fluc_parameterization}.

\subsection{Accidental Backgrounds}

During coincidence data taking, the rejection of gamma events is not perfect despite the PSD selection of neutron scatters with the LS detectors.
The accidental pileup of a neutron scatter in TPC and a gamma scatter in the LS detector is non-negligible since the rate of fixed-angle neutron scatters is low.
The S2-S1 distribution of this accidental pileup, included in the fit, is modeled to be the same as the band data.
The rate is left free for each set of coincidence data.

\subsection{Fitting Method}

A binned likelihood in log$_{10}$(S2/S1) versus S1 space, considering the data and MC simulation model described above, is used in the following statistical inference. The model parameter space is sampled by an affine invariant Markov Chain MC (MCMC)~\cite{emcee} algorithm, similarly to~\cite{xe100_tritium}.
The priors for $W$ and $p_{dpe}$ are taken from the NEST global fit~\cite{nest_v0.98} and a dedicated PMT double-PE measurement~\cite{dpe_2015}, respectively.
We also use the calibration data to provide priors for g$_1$, $G$, $\sigma_G$, $\sigma_{SPE}$, the electron extraction efficiency, S1 peak finding efficiency, and S2 trigger efficiency.
A GPU boosting is implemented in the MC simulation, similar to \citeref{anthony_pmt}, to be able to perform the fit within a reasonable time.

The fit is performed using all the available data (12 datasets) in this study.
There are, in total, 50 parameters that are varied in the fitting, 10 of which are for modeling the intrinsic light and charge signal response in LXe (Sec.~\ref{subsec:lxe_microphysics}), 16 are for modeling the signal reconstruction by detector (Sec.~\ref{subsec:detector_physics}), and 24 are for the NR rates and accidental pileup rates in each dataset.
Among all the parameters, $W$, $g_1$, $\sigma_{SPE}$, $p_{ext}$, $G$, $\sigma_G$, $p_{dpe}$, and 4 parameters for parameterizing the peak finding and trigger efficiencies are constrained with priors.
The rest of the fitting parameters are free.

\section{Results and Discussion}

\begin{figure*}[htp]
\centering
\includegraphics[width=0.8\textwidth]{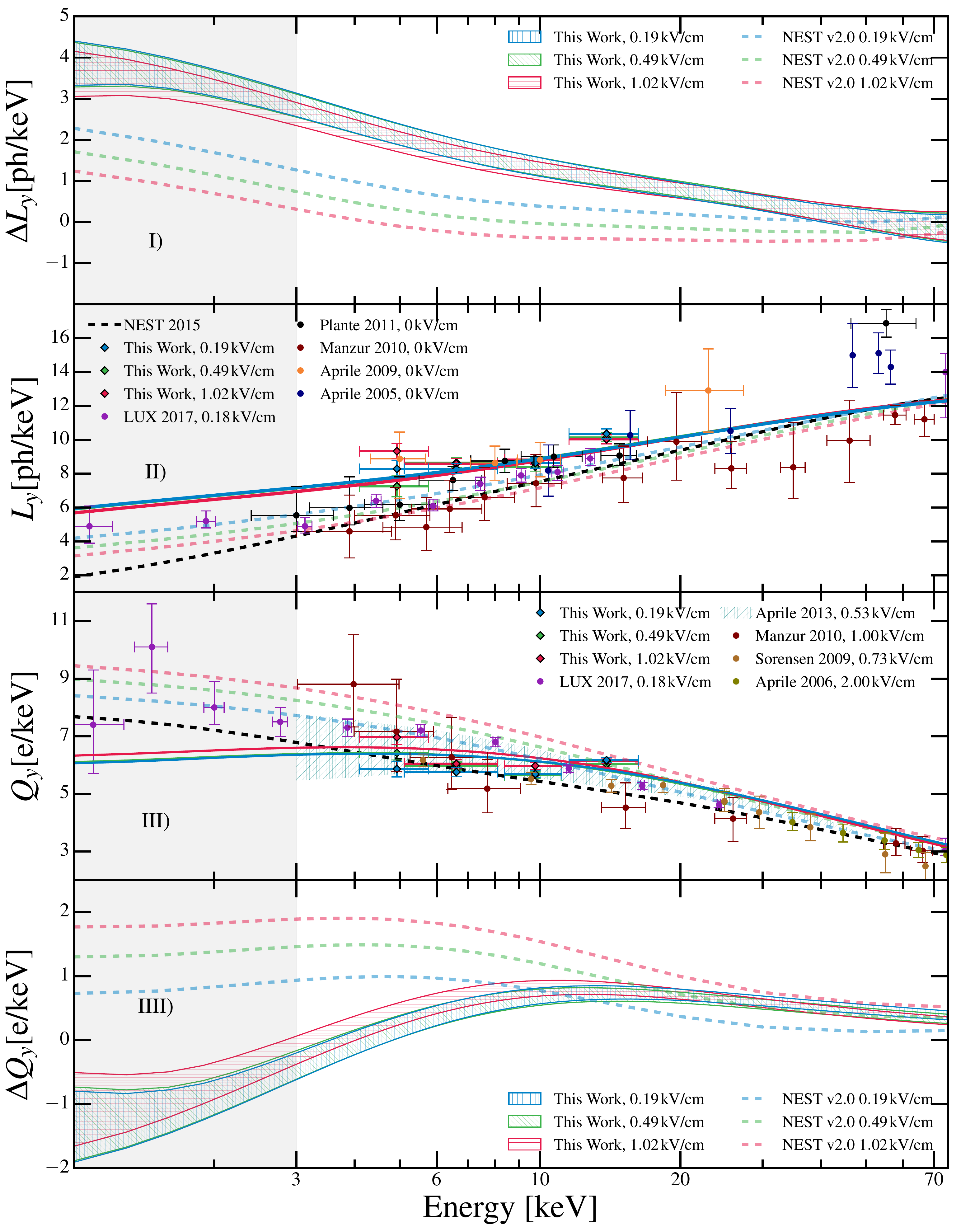}
\caption{
The comparison of the light and charge yield results from this work to previous measurements~\cite{aprile2005scintillation,prl2006,sorensen2009scintillation,aprile2009,manzur2010scintillation,horn2011nuclear,plante2011new, aprile2013response,lux_nr}.
Panel II) and III) show the light and charge yields, respectively.
The red, blue and green diamonds with error bars correspond to electric fields of 0.19, 0.49, and 1.02\,kV/cm, respectively, using the previous fixed-angle measurement interpretation as described in the text.
The red, blue and green solid lines are the results (medians of the posteriors) from the simultaneous fit of all available data with the MC simulation-based model.
The dashed lines serve as reference light and charge yields, which are the best-fit results from NEST v1.0\cite{nest_nr} at 490\,V/cm.
Panel I) and IIII) shows the deviations of the light and charge yield posteriors at the three fields to the reference NEST light and charge yields.
An energy cutoff of 3 keV (shown in gray) is chosen as this represents an approximately 10\% detection efficiency assuming a light yield of 5.5 photons/keV and charge yield of 7.5 electrons/keV.
Note that all measurements of light yield, with the exception of this work and of \citeref{lux_nr}, are extrapolations from $\mathcal{L}_{eff}$ using a light yield of 63 photons/keV for 122 keV electronic recoils.
For reference, light and charge yields from NEST v2.0~\cite{nest_v2.0} at 0.19, 0.49, 1.02\,kV/cm are shown in blue, green, and red dashed lines, respectively.
}
\label{fig:final_yields}
\end{figure*}

Figs.~\ref{fig:nerix_nr_best_fit_1}--\ref{fig:nerix_nr_best_fit_3} show the coincidence data compared to the MC simulations using best-fit parameters for all four angles and three electric fields.
\figref{fig:nerix_nr_best_fit_4} shows the band data comparison at the three electric fields.
No constrained parameter showed significant deviation from its prior in the fit.
From the fit results, the relative fluctuations from S1 noise, $\sigma_{s1}$/N$_{PE}$, are about 24\% and 19\% for S1 at 2\,PE and above 4\,PE, respectively.
The relative fluctuations from S2 noise,  $\sigma_{s2}$/(GN$_{ext}$), are about 33\%, 21\%, and 10\% for S2 at 500\,PE, 1000\,PE, and above 3000\,PE, respectively.
The goodness-of-fit (GoF) p-values of the data-model matching are calculated using the Gelman test~\cite{gof_gelman}.
The results are shown in Table~\ref{tab:GoFs_gelman}, with p-values larger than 1\% for most data at different fields.
The two low p-values of coincidence data with mean scattering energy of 4.9\,keV taken at 0.19\,kV/cm and band data taken at 1.02\,kV/cm are due to the anomalous events outside the most likely signal region (1$\sigma$ contour of the best-fit model, shown as the red solid line in Fig.~\ref{fig:nerix_nr_best_fit_1} to Fig.~\ref{fig:nerix_nr_best_fit_4}).
Excluding these events yields improved p-values of about 0.05 for the two datasets.

\begin{table}[htp]
\centering
\def\arraystretch{1.3}
\begin{tabular}{c||ccccc}
\hline\hline
\multirow{ 2}{*}{Field [V/cm]} & \multirow{ 2}{*}{Band data} & \multicolumn{4}{c}{Coincidence data} \\
& & 4.9\,keV & 6.6\,keV & 10.6\,keV & 14.0\,keV \\
\hline
190 &0.126 & 0.063 & 0.636 & 0.000 & 0.644 \\
490 & 0.478 & 0.293 & 0.606 & 0.712 & 0.704 \\
1020 & 0.004 & 0.012 & 0.461 & 0.784 & 0.181 \\
\hline\hline
\end{tabular}
\caption{
The estimated GoF p-values of the best-fit signal response model to all the coincidence and band data used.
The p-values are calcuated based on the Gelman method~\cite{gof_gelman}.
}
\label{tab:GoFs_gelman}
\end{table}

\begin{table}[htp]
\centering
\def\arraystretch{1.3}
\begin{tabular}{c||cccc}
\hline\hline
\multirow{ 2}{*}{Field [kV/cm]} & 4.9\,keV & 6.6\,keV & 10.6\,keV & 14.0\,keV  \\
& \multicolumn{4}{c}{L$_y$[ph/keV]} \\
\hline
0.19 & 8.3$\pm$0.5 & 8.3$\pm$0.3 & 8.6$\pm$0.3 & 10.4$\pm$0.3 \\
0.49 & 7.3$\pm$0.7 & 8.7$\pm$0.3 & 8.4$\pm$0.3 & 10.1$\pm$0.3 \\
1.02 & 9.3$\pm$0.5 & 8.6$\pm$0.3 & 8.8$\pm$0.3 & 10.0$\pm$0.3 \\
\hline
 & \multicolumn{4}{c}{Q$_y$[e/keV]} \\
\hline
0.19 & 5.9$\pm$0.3 & 5.8$\pm$0.1 & 5.7$\pm$0.1 & 6.2$\pm$0.1 \\
0.49 & 6.4$\pm$0.3 & 6.0$\pm$0.1 & 5.7$\pm$0.1 & 6.0$\pm$0.1 \\
1.02 & 7.0$\pm$0.3 & 6.0$\pm$0.1 & 6.0$\pm$0.1 & 6.1$\pm$0.1 \\
\hline\hline
\end{tabular}
\caption{
The scintillation and ionization yields derived using the traditional approach based on 1-D S1 and S2 spectra fitting, respectively.
}
\label{tab:py_qy_yields}
\end{table}

The systematic uncertainties induced by the uncertainties in positions of the neutron generator and LS detectors, and by the different binnings of the likelihood in log$_{10}$(S2/S1) versus S1 space were studied.
The total systematic uncertainty from these two factors are estimated to be less than 20\% of the statistical uncertainty, and are thus subdominant and not considered in the final parameter inference.


\figref{fig:final_yields} shows both the best-fits and credible regions for the light and charge yields at the different fields measured in this work, compared to recent measurements.  One striking feature of the yields is that this work found no statistically significant difference in the yields at the fields used in the nuclear recoil energies measured (about 3-70\,keV).  This supports the results of \citeref{prl2006, manzur2010scintillation} which only measured the effect above $\sim$45~keV.  Also of note is the slight disagreement in results when using a physical model versus the traditional model where a single light and charge yield (listed in Table~\ref{tab:py_qy_yields}), along with generic smearing terms, are used to fit the S1 and S2 spectra, separately.   In this work, the disagreement is likely due to the close proximity of the liquid scintillator detectors to the LXe detector, which broadens the expected energy spectrum, relative to \citeref{plante2011new}. Future fixed-angle measurements should consider this effect in their analysis and attempt to avoid this potential bias by utilizing a more physically motivated model.


With regards to the assumed model, we found a proportionality constant of $k = 0.188^{+0.008}_{-0.007}$, which is slightly larger than the value found in \citeref{lux_nr}  but is in agreement with the range of 0.1--0.2 from \citeref{lindhard1963integral}, and a quenching constant for Birk's saturation law of $\eta = 1.85^{+0.27}_{-0.26}$, which is in agreement with \citeref{lux_nr}.  Unlike \citeref{nest_nr}, our data don't show a significant energy dependence in the exciton-to-ion ratio.  However, we found a Thomas-Imel constant $\varsigma$ of $0.0087 \pm 0.0005$ and an exciton-to-ion ratio of $1.06 \pm 0.07$, both of which are in agreement with \citeref{nest_nr}.


\vspace{0.2cm}
\section{Acknowledgement}

We gratefully acknowledge support from the National Science Foundation for the XENON1T Dark Matter experiment at Columbia University (Award Numbers 1413495 and 1719286).



\begin{thebibliography}{37}
\expandafter\ifx\csname natexlab\endcsname\relax\def\natexlab#1{#1}\fi
\expandafter\ifx\csname bibnamefont\endcsname\relax
  \def\bibnamefont#1{#1}\fi
\expandafter\ifx\csname bibfnamefont\endcsname\relax
  \def\bibfnamefont#1{#1}\fi
\expandafter\ifx\csname citenamefont\endcsname\relax
  \def\citenamefont#1{#1}\fi
\expandafter\ifx\csname url\endcsname\relax
  \def\url#1{\texttt{#1}}\fi
\expandafter\ifx\csname urlprefix\endcsname\relax\def\urlprefix{URL }\fi
\providecommand{\bibinfo}[2]{#2}
\providecommand{\eprint}[2][]{\url{#2}}

\bibitem[{\citenamefont{Aprile et~al.}(2017{\natexlab{a}})}]{aprile2017first}
\bibinfo{author}{\bibfnamefont{E.}~\bibnamefont{Aprile}} \bibnamefont{et~al.}
  (\bibinfo{collaboration}{XENON Collaboration}), \bibinfo{journal}{Physical
  Review Letters} \textbf{\bibinfo{volume}{119}}, \bibinfo{pages}{181301}
  (\bibinfo{year}{2017}{\natexlab{a}}).

\bibitem[{\citenamefont{Cui et~al.}(2017)}]{cui2017dark}
\bibinfo{author}{\bibfnamefont{X.}~\bibnamefont{Cui}} \bibnamefont{et~al.}
  (\bibinfo{collaboration}{PandaX-II Collaboration}),
  \bibinfo{journal}{Physical Review Letters} \textbf{\bibinfo{volume}{119}},
  \bibinfo{pages}{181302} (\bibinfo{year}{2017}).

\bibitem[{\citenamefont{Akerib
  et~al.}(2016{\natexlab{a}})}]{akerib2016improved}
\bibinfo{author}{\bibfnamefont{D.}~\bibnamefont{Akerib}} \bibnamefont{et~al.}
  (\bibinfo{collaboration}{LUX Collaboration}), \bibinfo{journal}{Physical
  Review Letters} \textbf{\bibinfo{volume}{116}}, \bibinfo{pages}{161301}
  (\bibinfo{year}{2016}{\natexlab{a}}).

\bibitem[{\citenamefont{Aprile et~al.}(2016)}]{aprile2016xenon100}
\bibinfo{author}{\bibfnamefont{E.}~\bibnamefont{Aprile}} \bibnamefont{et~al.}
  (\bibinfo{collaboration}{XENON Collaboration}), \bibinfo{journal}{Physical
  Review D} \textbf{\bibinfo{volume}{94}}, \bibinfo{pages}{122001}
  (\bibinfo{year}{2016}).

\bibitem[{\citenamefont{Aprile et~al.}(2017{\natexlab{b}})}]{aprile2017xenon1t}
\bibinfo{author}{\bibfnamefont{E.}~\bibnamefont{Aprile}} \bibnamefont{et~al.}
  (\bibinfo{collaboration}{XENON Collaboration}), \bibinfo{journal}{The
  European Physical Journal C} \textbf{\bibinfo{volume}{77}},
  \bibinfo{pages}{881} (\bibinfo{year}{2017}{\natexlab{b}}).

\bibitem[{\citenamefont{Aprile et~al.}(2005)\citenamefont{Aprile, Giboni,
  Majewski, Ni, Yamashita, Hasty, Manzur, and
  McKinsey}}]{aprile2005scintillation}
\bibinfo{author}{\bibfnamefont{E.}~\bibnamefont{Aprile}},
  \bibinfo{author}{\bibfnamefont{K.}~\bibnamefont{Giboni}},
  \bibinfo{author}{\bibfnamefont{P.}~\bibnamefont{Majewski}},
  \bibinfo{author}{\bibfnamefont{K.}~\bibnamefont{Ni}},
  \bibinfo{author}{\bibfnamefont{M.}~\bibnamefont{Yamashita}},
  \bibinfo{author}{\bibfnamefont{R.}~\bibnamefont{Hasty}},
  \bibinfo{author}{\bibfnamefont{A.}~\bibnamefont{Manzur}}, \bibnamefont{and}
  \bibinfo{author}{\bibfnamefont{D.}~\bibnamefont{McKinsey}},
  \bibinfo{journal}{Physical Review D} \textbf{\bibinfo{volume}{72}},
  \bibinfo{pages}{072006} (\bibinfo{year}{2005}).

\bibitem[{\citenamefont{Aprile et~al.}(2006)\citenamefont{Aprile, Dahl,
  de~Viveiros, Gaitskell, Giboni, Kwong, Majewski, Ni, Shutt, and
  Yamashita}}]{prl2006}
\bibinfo{author}{\bibfnamefont{E.}~\bibnamefont{Aprile}},
  \bibinfo{author}{\bibfnamefont{C.}~\bibnamefont{Dahl}},
  \bibinfo{author}{\bibfnamefont{L.}~\bibnamefont{de~Viveiros}},
  \bibinfo{author}{\bibfnamefont{R.}~\bibnamefont{Gaitskell}},
  \bibinfo{author}{\bibfnamefont{K.-L.} \bibnamefont{Giboni}},
  \bibinfo{author}{\bibfnamefont{J.}~\bibnamefont{Kwong}},
  \bibinfo{author}{\bibfnamefont{P.}~\bibnamefont{Majewski}},
  \bibinfo{author}{\bibfnamefont{K.}~\bibnamefont{Ni}},
  \bibinfo{author}{\bibfnamefont{T.}~\bibnamefont{Shutt}}, \bibnamefont{and}
  \bibinfo{author}{\bibfnamefont{M.}~\bibnamefont{Yamashita}},
  \bibinfo{journal}{Physical Review Letters} \textbf{\bibinfo{volume}{97}},
  \bibinfo{pages}{081302} (\bibinfo{year}{2006}).

\bibitem[{\citenamefont{Sorensen et~al.}(2009)}]{sorensen2009scintillation}
\bibinfo{author}{\bibfnamefont{P.}~\bibnamefont{Sorensen}} \bibnamefont{et~al.}
  (\bibinfo{collaboration}{XENON10 Collaboration}), \bibinfo{journal}{Nuclear
  Instruments and Methods in Physics Research Section A: Accelerators,
  Spectrometers, Detectors and Associated Equipment}
  \textbf{\bibinfo{volume}{601}}, \bibinfo{pages}{339} (\bibinfo{year}{2009}).

\bibitem[{\citenamefont{Aprile et~al.}(2009)\citenamefont{Aprile, Baudis, Choi,
  Giboni, Lim, Manalaysay, Monzani, Plante, Santorelli, and
  Yamashita}}]{aprile2009}
\bibinfo{author}{\bibfnamefont{E.}~\bibnamefont{Aprile}},
  \bibinfo{author}{\bibfnamefont{L.}~\bibnamefont{Baudis}},
  \bibinfo{author}{\bibfnamefont{B.}~\bibnamefont{Choi}},
  \bibinfo{author}{\bibfnamefont{K.}~\bibnamefont{Giboni}},
  \bibinfo{author}{\bibfnamefont{K.}~\bibnamefont{Lim}},
  \bibinfo{author}{\bibfnamefont{A.}~\bibnamefont{Manalaysay}},
  \bibinfo{author}{\bibfnamefont{M.}~\bibnamefont{Monzani}},
  \bibinfo{author}{\bibfnamefont{G.}~\bibnamefont{Plante}},
  \bibinfo{author}{\bibfnamefont{R.}~\bibnamefont{Santorelli}},
  \bibnamefont{and}
  \bibinfo{author}{\bibfnamefont{M.}~\bibnamefont{Yamashita}},
  \bibinfo{journal}{Physical Review C} \textbf{\bibinfo{volume}{79}},
  \bibinfo{pages}{045807} (\bibinfo{year}{2009}).

\bibitem[{\citenamefont{Manzur et~al.}(2010)\citenamefont{Manzur, Curioni,
  Kastens, McKinsey, Ni, and Wongjirad}}]{manzur2010scintillation}
\bibinfo{author}{\bibfnamefont{A.}~\bibnamefont{Manzur}},
  \bibinfo{author}{\bibfnamefont{A.}~\bibnamefont{Curioni}},
  \bibinfo{author}{\bibfnamefont{L.}~\bibnamefont{Kastens}},
  \bibinfo{author}{\bibfnamefont{D.}~\bibnamefont{McKinsey}},
  \bibinfo{author}{\bibfnamefont{K.}~\bibnamefont{Ni}}, \bibnamefont{and}
  \bibinfo{author}{\bibfnamefont{T.}~\bibnamefont{Wongjirad}},
  \bibinfo{journal}{Physical Review C} \textbf{\bibinfo{volume}{81}},
  \bibinfo{pages}{025808} (\bibinfo{year}{2010}).

\bibitem[{\citenamefont{Horn et~al.}(2011)}]{horn2011nuclear}
\bibinfo{author}{\bibfnamefont{M.}~\bibnamefont{Horn}} \bibnamefont{et~al.}
  (\bibinfo{collaboration}{ZEPLIN-III Collaboration}),
  \bibinfo{journal}{Physics Letters B} \textbf{\bibinfo{volume}{705}},
  \bibinfo{pages}{471} (\bibinfo{year}{2011}).

\bibitem[{\citenamefont{Plante et~al.}(2011)\citenamefont{Plante, Aprile,
  Budnik, Choi, Giboni, Goetzke, Lang, Lim, and Fernandez}}]{plante2011new}
\bibinfo{author}{\bibfnamefont{G.}~\bibnamefont{Plante}},
  \bibinfo{author}{\bibfnamefont{E.}~\bibnamefont{Aprile}},
  \bibinfo{author}{\bibfnamefont{R.}~\bibnamefont{Budnik}},
  \bibinfo{author}{\bibfnamefont{B.}~\bibnamefont{Choi}},
  \bibinfo{author}{\bibfnamefont{K.-L.} \bibnamefont{Giboni}},
  \bibinfo{author}{\bibfnamefont{L.}~\bibnamefont{Goetzke}},
  \bibinfo{author}{\bibfnamefont{R.}~\bibnamefont{Lang}},
  \bibinfo{author}{\bibfnamefont{K.}~\bibnamefont{Lim}}, \bibnamefont{and}
  \bibinfo{author}{\bibfnamefont{A.~M.} \bibnamefont{Fernandez}},
  \bibinfo{journal}{Physical Review C} \textbf{\bibinfo{volume}{84}},
  \bibinfo{pages}{045805} (\bibinfo{year}{2011}).

\bibitem[{\citenamefont{Aprile et~al.}(2013{\natexlab{a}})}]{xe100_nr}
\bibinfo{author}{\bibfnamefont{E.}~\bibnamefont{Aprile}} \bibnamefont{et~al.}
  (\bibinfo{collaboration}{XENON100 Collaboration}), \bibinfo{journal}{Physical
  Review D} \textbf{\bibinfo{volume}{88}}, \bibinfo{pages}{012006}
  (\bibinfo{year}{2013}{\natexlab{a}}).

\bibitem[{\citenamefont{Akerib et~al.}(2016{\natexlab{b}})}]{lux_nr}
\bibinfo{author}{\bibfnamefont{D.}~\bibnamefont{Akerib}} \bibnamefont{et~al.}
  (\bibinfo{collaboration}{LUX Collaboration}), \bibinfo{journal}{arXiv
  preprint arXiv:1608.05381}  (\bibinfo{year}{2016}{\natexlab{b}}).

\bibitem[{\citenamefont{Lenardo et~al.}(2015)\citenamefont{Lenardo, Kazkaz,
  Manalaysay, Mock, Szydagis, and Tripathi}}]{nest_nr}
\bibinfo{author}{\bibfnamefont{B.}~\bibnamefont{Lenardo}},
  \bibinfo{author}{\bibfnamefont{K.}~\bibnamefont{Kazkaz}},
  \bibinfo{author}{\bibfnamefont{A.}~\bibnamefont{Manalaysay}},
  \bibinfo{author}{\bibfnamefont{J.}~\bibnamefont{Mock}},
  \bibinfo{author}{\bibfnamefont{M.}~\bibnamefont{Szydagis}}, \bibnamefont{and}
  \bibinfo{author}{\bibfnamefont{M.}~\bibnamefont{Tripathi}},
  \bibinfo{journal}{IEEE Transactions on Nuclear Science}
  \textbf{\bibinfo{volume}{62}}, \bibinfo{pages}{3387} (\bibinfo{year}{2015}).

\bibitem[{\citenamefont{Aprile et~al.}(2018{\natexlab{a}})}]{xe1t_sr1}
\bibinfo{author}{\bibfnamefont{E.}~\bibnamefont{Aprile}} \bibnamefont{et~al.}
  (\bibinfo{collaboration}{XENON Collaboration}), \bibinfo{journal}{arXiv
  preprint arXiv:1805.12562}  (\bibinfo{year}{2018}{\natexlab{a}}).

\bibitem[{\citenamefont{Lindhard et~al.}(1963)\citenamefont{Lindhard, Nielsen,
  Scharff, and Thomsen}}]{lindhard1963integral}
\bibinfo{author}{\bibfnamefont{J.}~\bibnamefont{Lindhard}},
  \bibinfo{author}{\bibfnamefont{V.}~\bibnamefont{Nielsen}},
  \bibinfo{author}{\bibfnamefont{M.}~\bibnamefont{Scharff}}, \bibnamefont{and}
  \bibinfo{author}{\bibfnamefont{P.}~\bibnamefont{Thomsen}},
  \bibinfo{journal}{Mat. Fys. Medd. Dan. Vid. Selsk}
  \textbf{\bibinfo{volume}{33}}, \bibinfo{pages}{1} (\bibinfo{year}{1963}).

\bibitem[{\citenamefont{Goetzke et~al.}(2017)\citenamefont{Goetzke, Aprile,
  Anthony, Plante, and Weber}}]{goetzke2016measurement}
\bibinfo{author}{\bibfnamefont{L.~W.} \bibnamefont{Goetzke}},
  \bibinfo{author}{\bibfnamefont{E.}~\bibnamefont{Aprile}},
  \bibinfo{author}{\bibfnamefont{M.}~\bibnamefont{Anthony}},
  \bibinfo{author}{\bibfnamefont{G.}~\bibnamefont{Plante}}, \bibnamefont{and}
  \bibinfo{author}{\bibfnamefont{M.}~\bibnamefont{Weber}},
  \bibinfo{journal}{Phys. Rev. D} \textbf{\bibinfo{volume}{96}},
  \bibinfo{pages}{103007} (\bibinfo{year}{2017}).

\bibitem[{\citenamefont{Thomas and Imel}(1987)}]{ti_recombination}
\bibinfo{author}{\bibfnamefont{J.}~\bibnamefont{Thomas}} \bibnamefont{and}
  \bibinfo{author}{\bibfnamefont{D.}~\bibnamefont{Imel}},
  \bibinfo{journal}{Physical Review A} \textbf{\bibinfo{volume}{36}},
  \bibinfo{pages}{614} (\bibinfo{year}{1987}).

\bibitem[{\citenamefont{Aprile and Doke}(2010)}]{aprile2010liquid}
\bibinfo{author}{\bibfnamefont{E.}~\bibnamefont{Aprile}} \bibnamefont{and}
  \bibinfo{author}{\bibfnamefont{T.}~\bibnamefont{Doke}},
  \bibinfo{journal}{Reviews of Modern Physics} \textbf{\bibinfo{volume}{82}},
  \bibinfo{pages}{2053} (\bibinfo{year}{2010}).

\bibitem[{\citenamefont{Goetzke}(2015)}]{luke_thesis}
\bibinfo{author}{\bibfnamefont{L.}~\bibnamefont{Goetzke}}, \bibinfo{type}{{PhD}
  dissertation}, \bibinfo{school}{Columbia University} (\bibinfo{year}{2015}).

\bibitem[{\citenamefont{{COMSOL Multiphysics \circledR v.5.2. COMSOL
  AB}}(2015)}]{COMSOL}
\bibinfo{author}{\bibnamefont{{COMSOL Multiphysics \circledR v.5.2. COMSOL
  AB}}}, \emph{\bibinfo{title}{AC/DC Module User's Guide}},
  \bibinfo{address}{Stockholm, Sweden.} (\bibinfo{year}{2015}).

\bibitem[{\citenamefont{{Eljen Technology}}(2016)}]{ej301_manual}
\bibinfo{author}{\bibnamefont{{Eljen Technology}}},
  \emph{\bibinfo{title}{Neutron/Gamma PSD Liquid Scintillator EJ-301, EJ-309}},
  \bibinfo{address}{1300 W. Broadway, Sweetwater, TX 79556}
  (\bibinfo{year}{2016}).

\bibitem[{\citenamefont{Agostinelli et~al.}(2003)}]{geant4}
\bibinfo{author}{\bibfnamefont{S.}~\bibnamefont{Agostinelli}}
  \bibnamefont{et~al.} (\bibinfo{collaboration}{Geant4 Collaboration}),
  \bibinfo{journal}{Nuclear instruments and methods in physics research section
  A: Accelerators, Spectrometers, Detectors and Associated Equipment}
  \textbf{\bibinfo{volume}{506}}, \bibinfo{pages}{250} (\bibinfo{year}{2003}).

\bibitem[{\citenamefont{Aprile et~al.}(2014)\citenamefont{Aprile, Alfonsi,
  Arisaka, Arneodo, Balan, Baudis, Behrens, Beltrame, Bokeloh, Brown
  et~al.}}]{aprile2014analysis}
\bibinfo{author}{\bibfnamefont{E.}~\bibnamefont{Aprile}},
  \bibinfo{author}{\bibfnamefont{M.}~\bibnamefont{Alfonsi}},
  \bibinfo{author}{\bibfnamefont{K.}~\bibnamefont{Arisaka}},
  \bibinfo{author}{\bibfnamefont{F.}~\bibnamefont{Arneodo}},
  \bibinfo{author}{\bibfnamefont{C.}~\bibnamefont{Balan}},
  \bibinfo{author}{\bibfnamefont{L.}~\bibnamefont{Baudis}},
  \bibinfo{author}{\bibfnamefont{A.}~\bibnamefont{Behrens}},
  \bibinfo{author}{\bibfnamefont{P.}~\bibnamefont{Beltrame}},
  \bibinfo{author}{\bibfnamefont{K.}~\bibnamefont{Bokeloh}},
  \bibinfo{author}{\bibfnamefont{E.}~\bibnamefont{Brown}},
  \bibnamefont{et~al.}, \bibinfo{journal}{Astroparticle Physics}
  \textbf{\bibinfo{volume}{54}}, \bibinfo{pages}{11} (\bibinfo{year}{2014}).

\bibitem[{\citenamefont{Aprile et~al.}(2010)}]{xe1t_instr}
\bibinfo{author}{\bibfnamefont{E.}~\bibnamefont{Aprile}} \bibnamefont{et~al.}
  (\bibinfo{collaboration}{XENON Collaboration}), \bibinfo{journal}{The
  European Physical Journal C} \textbf{\bibinfo{volume}{77}},
  \bibinfo{pages}{881} (\bibinfo{year}{2010}).

\bibitem[{\citenamefont{Doke et~al.}(2002)\citenamefont{Doke, Hitachi, Kikuchi,
  Masuda, Okada, and Shibamura}}]{doke_ac}
\bibinfo{author}{\bibfnamefont{T.}~\bibnamefont{Doke}},
  \bibinfo{author}{\bibfnamefont{A.}~\bibnamefont{Hitachi}},
  \bibinfo{author}{\bibfnamefont{J.}~\bibnamefont{Kikuchi}},
  \bibinfo{author}{\bibfnamefont{K.}~\bibnamefont{Masuda}},
  \bibinfo{author}{\bibfnamefont{H.}~\bibnamefont{Okada}}, \bibnamefont{and}
  \bibinfo{author}{\bibfnamefont{E.}~\bibnamefont{Shibamura}},
  \bibinfo{journal}{Japanese journal of applied physics}
  \textbf{\bibinfo{volume}{41}}, \bibinfo{pages}{1538} (\bibinfo{year}{2002}).

\bibitem[{\citenamefont{Faham et~al.}(2015)\citenamefont{Faham, Gehman, Currie,
  Dobi, Sorensen, and Gaitskell}}]{dpe_2015}
\bibinfo{author}{\bibfnamefont{C.}~\bibnamefont{Faham}},
  \bibinfo{author}{\bibfnamefont{V.}~\bibnamefont{Gehman}},
  \bibinfo{author}{\bibfnamefont{A.}~\bibnamefont{Currie}},
  \bibinfo{author}{\bibfnamefont{A.}~\bibnamefont{Dobi}},
  \bibinfo{author}{\bibfnamefont{P.}~\bibnamefont{Sorensen}}, \bibnamefont{and}
  \bibinfo{author}{\bibfnamefont{R.}~\bibnamefont{Gaitskell}},
  \bibinfo{journal}{Journal of Instrumentation} \textbf{\bibinfo{volume}{10}},
  \bibinfo{pages}{P09010} (\bibinfo{year}{2015}).

\bibitem[{\citenamefont{Hitachi et~al.}(1983)\citenamefont{Hitachi, Takahashi,
  Funayama, Masuda, Kikuchi, and Doke}}]{s1_shape}
\bibinfo{author}{\bibfnamefont{A.}~\bibnamefont{Hitachi}},
  \bibinfo{author}{\bibfnamefont{T.}~\bibnamefont{Takahashi}},
  \bibinfo{author}{\bibfnamefont{N.}~\bibnamefont{Funayama}},
  \bibinfo{author}{\bibfnamefont{K.}~\bibnamefont{Masuda}},
  \bibinfo{author}{\bibfnamefont{J.}~\bibnamefont{Kikuchi}}, \bibnamefont{and}
  \bibinfo{author}{\bibfnamefont{T.}~\bibnamefont{Doke}},
  \bibinfo{journal}{Physical Review B} \textbf{\bibinfo{volume}{27}},
  \bibinfo{pages}{5279} (\bibinfo{year}{1983}).

\bibitem[{\citenamefont{Aprile et~al.}(2018{\natexlab{b}})}]{xe100_tritium}
\bibinfo{author}{\bibfnamefont{E.}~\bibnamefont{Aprile}} \bibnamefont{et~al.}
  (\bibinfo{collaboration}{XENON Collaboration}), \bibinfo{journal}{Physical
  Review D} \textbf{\bibinfo{volume}{97}}, \bibinfo{pages}{092007}
  (\bibinfo{year}{2018}{\natexlab{b}}).

\bibitem[{\citenamefont{Szydagis et~al.}(2011)\citenamefont{Szydagis, Barry,
  Kazkaz, Mock, Stolp, Sweany, Tripathi, Uvarov, Walsh, and
  Woods}}]{nest_v0.98}
\bibinfo{author}{\bibfnamefont{M.}~\bibnamefont{Szydagis}},
  \bibinfo{author}{\bibfnamefont{N.}~\bibnamefont{Barry}},
  \bibinfo{author}{\bibfnamefont{K.}~\bibnamefont{Kazkaz}},
  \bibinfo{author}{\bibfnamefont{J.}~\bibnamefont{Mock}},
  \bibinfo{author}{\bibfnamefont{D.}~\bibnamefont{Stolp}},
  \bibinfo{author}{\bibfnamefont{M.}~\bibnamefont{Sweany}},
  \bibinfo{author}{\bibfnamefont{M.}~\bibnamefont{Tripathi}},
  \bibinfo{author}{\bibfnamefont{S.}~\bibnamefont{Uvarov}},
  \bibinfo{author}{\bibfnamefont{N.}~\bibnamefont{Walsh}}, \bibnamefont{and}
  \bibinfo{author}{\bibfnamefont{M.}~\bibnamefont{Woods}},
  \bibinfo{journal}{Journal of Instrumentation} \textbf{\bibinfo{volume}{6}},
  \bibinfo{pages}{P10002} (\bibinfo{year}{2011}).

\bibitem[{\citenamefont{Bezrukov et~al.}(2011)\citenamefont{Bezrukov,
  Kahlhoefer, and Lindner}}]{bezrukov2011interplay}
\bibinfo{author}{\bibfnamefont{F.}~\bibnamefont{Bezrukov}},
  \bibinfo{author}{\bibfnamefont{F.}~\bibnamefont{Kahlhoefer}},
  \bibnamefont{and} \bibinfo{author}{\bibfnamefont{M.}~\bibnamefont{Lindner}},
  \bibinfo{journal}{Astroparticle Physics} \textbf{\bibinfo{volume}{35}},
  \bibinfo{pages}{119} (\bibinfo{year}{2011}).

\bibitem[{\citenamefont{Foreman-Mackey
  et~al.}(2013)\citenamefont{Foreman-Mackey, Hogg, Lang, and Goodman}}]{emcee}
\bibinfo{author}{\bibfnamefont{D.}~\bibnamefont{Foreman-Mackey}},
  \bibinfo{author}{\bibfnamefont{D.~W.} \bibnamefont{Hogg}},
  \bibinfo{author}{\bibfnamefont{D.}~\bibnamefont{Lang}}, \bibnamefont{and}
  \bibinfo{author}{\bibfnamefont{J.}~\bibnamefont{Goodman}},
  \bibinfo{journal}{Publications of the Astronomical Society of the Pacific}
  \textbf{\bibinfo{volume}{125}}, \bibinfo{pages}{306} (\bibinfo{year}{2013}).

\bibitem[{\citenamefont{Anthony et~al.}(2018)\citenamefont{Anthony, Aprile,
  Grandi, Lin, and Saldanha}}]{anthony_pmt}
\bibinfo{author}{\bibfnamefont{M.}~\bibnamefont{Anthony}},
  \bibinfo{author}{\bibfnamefont{E.}~\bibnamefont{Aprile}},
  \bibinfo{author}{\bibfnamefont{L.}~\bibnamefont{Grandi}},
  \bibinfo{author}{\bibfnamefont{Q.}~\bibnamefont{Lin}}, \bibnamefont{and}
  \bibinfo{author}{\bibfnamefont{R.}~\bibnamefont{Saldanha}},
  \bibinfo{journal}{Journal of Instrumentation} \textbf{\bibinfo{volume}{13}},
  \bibinfo{pages}{T02011} (\bibinfo{year}{2018}).

\bibitem[{\citenamefont{Aprile
  et~al.}(2013{\natexlab{b}})}]{aprile2013response}
\bibinfo{author}{\bibfnamefont{E.}~\bibnamefont{Aprile}} \bibnamefont{et~al.}
  (\bibinfo{collaboration}{XENON Collaboration}), \bibinfo{journal}{Physical
  Review D} \textbf{\bibinfo{volume}{88}}, \bibinfo{pages}{012006}
  (\bibinfo{year}{2013}{\natexlab{b}}).

\bibitem[{\citenamefont{Szydagis et~al.}(2018)\citenamefont{Szydagis, Balajthy,
  Brodsky, Cutter, Huang, Kozlova, Lenardo, Manalaysay, McKinsey, Mooney
  et~al.}}]{nest_v2.0}
\bibinfo{author}{\bibfnamefont{M.}~\bibnamefont{Szydagis}},
  \bibinfo{author}{\bibfnamefont{J.}~\bibnamefont{Balajthy}},
  \bibinfo{author}{\bibfnamefont{J.}~\bibnamefont{Brodsky}},
  \bibinfo{author}{\bibfnamefont{J.}~\bibnamefont{Cutter}},
  \bibinfo{author}{\bibfnamefont{J.}~\bibnamefont{Huang}},
  \bibinfo{author}{\bibfnamefont{E.}~\bibnamefont{Kozlova}},
  \bibinfo{author}{\bibfnamefont{B.}~\bibnamefont{Lenardo}},
  \bibinfo{author}{\bibfnamefont{A.}~\bibnamefont{Manalaysay}},
  \bibinfo{author}{\bibfnamefont{D.}~\bibnamefont{McKinsey}},
  \bibinfo{author}{\bibfnamefont{M.}~\bibnamefont{Mooney}},
  \bibnamefont{et~al.}, \emph{\bibinfo{title}{Noble element simulation
  technique v2.0}} (\bibinfo{year}{2018}),
  \urlprefix\url{https://doi.org/10.5281/zenodo.1314669}.

\bibitem[{\citenamefont{Gelman et~al.}(1996)\citenamefont{Gelman, Meng, and
  Stern}}]{gof_gelman}
\bibinfo{author}{\bibfnamefont{A.}~\bibnamefont{Gelman}},
  \bibinfo{author}{\bibfnamefont{X.-L.} \bibnamefont{Meng}}, \bibnamefont{and}
  \bibinfo{author}{\bibfnamefont{H.}~\bibnamefont{Stern}},
  \bibinfo{journal}{Statistica sinica} pp. \bibinfo{pages}{733--760}
  (\bibinfo{year}{1996}).

\end{thebibliography}

\end{document}